\documentclass[aps,pre,showpacs,superscriptaddress,longbibliography,twocolumn,floatfix]{revtex4-1}
\usepackage{graphicx}
\usepackage{amsmath}
\usepackage{amssymb}
\usepackage{mathtools}
\usepackage{makeidx}
\usepackage{amsfonts}
\usepackage{bm}

\begin{document}

\title{Possible instability of one-step replica symmetry breaking in $p$-spin Ising models outside mean-field theory}
\author{J.\ Yeo}
\affiliation{Department of Physics, Konkuk University, Seoul 05029, Korea}
\author{M.\ A.\ Moore}
\affiliation{Department of Physics and Astronomy, University of Manchester,
Manchester M13 9PL, United Kingdom.}

\date{\today}

\begin{abstract}
The fully-connected Ising $p$-spin model has for  $p >2$ a discontinuous  phase transition from the paramagnetic phase to a stable state with one-step replica symmetry breaking (1RSB). However, simulations in three dimension do not  look like these mean-field results and have features more like those which would arise with full replica symmetry breaking (FRSB).  To help understand how this might come about we have studied in the fully connected $p$-spin model the state of two-step replica symmetry breaking (2RSB). It has a free energy degenerate  with that of 1RSB, but the weight of the additional peak in $P(q)$ vanishes.  We expect that the state with full replica symmetry breaking (FRSB) is also degenerate with that of 1RSB. We suggest that finite size effects will give a non-vanishing weight to the FRSB features, as also  will fluctuations about the mean-field solution. Our conclusion is that outside the fully connected model in the thermodynamic limit, FRSB is to be expected rather than 1RSB.

\end{abstract}



\maketitle

\section{Introduction}

The historical importance of $p$-spin models \cite{gardner:85} was that it was investigations of their properties which led to the random first order transition theory (RFOT) of structural glasses \cite{kirkpatrick:89,Lubchenko:07,kirkpatrick:15,dzero:09,biroli:09, cavagna:09}. In this paper we shall once again study the fully-connected $p$-spin glass model described by the Hamiltonian \cite{gardner:85}
\begin{equation}
 H=-\sum_{1\le i_1<\cdots<i_p\le N} J_{i_1,i_2,\cdots,i_p}S_{i_1}S_{i_2}\cdots S_{i_p},
\label{Ham}
\end{equation}
where $S_i$ is the Ising spin at site $i$, $i,\cdots,N$, and the bonds $J_{i_1,i_2,\cdots,i_p}$ are independent random variables satisfying 
the Gaussian distribution with zero mean and variance $p!/(2N^{p-1})$. At high temperatures, there is a replica symmetric (RS) paramagnetic phase. To characterise the nature of a glass phase it is useful to study the distribution $P(q)$ of overlaps $q$, between two copies $A$ and $B$ of the system, where 
\begin{equation}
q = \frac{1}{N} \sum_i  S_i^A S_i^B,
\label{twocopyq}
\end{equation} 
and $P(q)$ is defined as
\begin{equation}
P(q)=\overline{P_J(q)}.
\end{equation}
 The thermal averaging is done with the Hamiltonians of the two copies, $A$ and $B$, and the overline denotes an average over the bonds. 
In the paramagnetic (high-temperature) phase of the Hamiltonian of Eq.~(\ref{Ham}),
\begin{equation}
P(q)=\delta(q-q_0),
\end{equation}
where  $q_0=0$. 
Much of the work on the $p$-spin model has focussed on the dynamical transition at a temperature $T_d$, the temperature below which metastable states exist separated by barriers of order $N$. This has connections with the glass transition $T_g$ of structural glasses.
At a lower temperature $T^c_{\rm 1RSB}$  there is a discontinuous phase transition to a state with one step replica symmetry breaking (1RSB). The distribution of overlaps in the low temperature phase becomes
\begin{equation}
P(q)=m \delta(q-q_0)+(1-m)\delta(q-q_1),
\label{Pq1RSB}
\end{equation}
where $0  \le m \le 1$ and $m$ equals unity at $T^c_{\rm 1RSB}$. The order parameter $q_1$ is finite at the transition, so the transition is discontinuous, but there is no latent heat at the transition. In the RFOT approach to structural glasses $T^c_{\rm 1RSB}$ becomes the Kauzmann transition temperature $T_K$ \cite{kauzmann:48}, and for temperatures below $T_K$ the phase with 1RSB is the ideal glass phase. This paper is about the properties of the $p$-spin system at temperatures less than  $T^c_{\rm 1RSB}$ or structural glasses at temperatures below $T_K$.  Until recently,
behavior at temperatures  below $T_g$  would have been  of only
academic interest,  but the use  of the swap algorithm  in simulations \cite{ediger:16}
and  the  creation of  ultrastable  glasses  by vapor  deposition  has
changed that situation \cite{ediger:16,Royall:18}. 

Gardner  pointed out  \cite{gardner:85}  that  in the  fully
  connected  $p$-spin  model    there is  a continuous  transition (now  called the
  Gardner transition), at  a temperature  $T_G$   below
  $T^c_{\rm 1RSB}$,   from the state with  1RSB   to a state  with full
  replica  symmetry  breaking (FRSB) \cite{gross:85}.  There has been  recently much  interest  in   Gardner
  transitions  in structural  glasses, which is reviewed  in
  Ref.~\cite{berthier:19}.   

The chief purpose of this paper is to question whether the state which exists between $T_G < T <T^c_{\rm 1RSB}$ really does have 1RSB symmetry breaking. We shall argue that it probably is much more like a state with FRSB. We reached this conclusion from the calculation done in Sec.~\ref{2RSB}, where we have gone beyond 1RSB to 2RSB. A 2RSB calculation has already been done for the $p$-spin model by Montanari and Ricci-Tersenghi \cite{montanari:03}, but it was mostly focussed on behavior at $T = 0$. In a 2RSB calculation $P(q)$ has the form
\begin{equation}
P(q)= m_1\delta(q-q_0)+(m_2-m_1) \delta(q-q_1)+(1-m_2)\delta(q-q_2),
\label{Pq2RSB}
\end{equation}
where the weights $1 \ge m_2 \ge m_1 \ge 0$.

 We shall find in Sec. \ref{2RSB} that there are two types of solution for the equations which determine  $m_1,m_2, q_0,q_1, q_2$ for the fully-connected $p$-spin model in the limit $N \to \infty$. There is a solution, which we refer to as the type-II 2RSB solution, which has a higher free energy than the 1RSB solution at temperature $T < T_G$ (which in the world of replicas makes it more stable than the 1RSB solution); it is, however, just the 2RSB approximation to the state of FRSB which is expected for $T < T_G$. It has an extension to the region $T > T_G$, but there it has a lower free energy than that of the state of 1RSB.

The other solution will be referred to as the  type-I 2RSB solution. It  has $m_2=m_1 =m$, where $m$ is that of the 1RSB solution of Eq.~(\ref{Pq1RSB}). The free energy of this solution is exactly degenerate with that of the 1RSB state, and because $m_2=m_1$ its $P(q)$ is also identical with that of the 1RSB state: There is no weight in the delta function at $q= q_1$. One would expect that similar states will exist at the $K$ level of replica symmetry breaking, KRSB, all the way up to FRSB. We believe that if one goes beyond the leading term as $N \to \infty$ to include the fluctuations around the mean-field limit then the degeneracy is lifted and the system goes to a state with FRSB, by having (say) $m_2-m_1$ no longer zero but of order of a small quantity which  vanishes as $N \to \infty$.

We can provide two arguments that 1RSB has to be replaced by FRSB. In fact we believe that there is already evidence that finite $N$ effects do modify 1RSB order to FRSB order in the fully-connected model. Billoire et al. \cite{billoire:05} studied $P(q)$ of the $p=3$ fully-connected model for values of $N$ up to $192$. Their $P(q)$ looks remarkably similar to the finite size form predicted by Mottishaw and Derrida \cite{mottishaw:18} for the generalized random energy model (GREM) when FRSB is present. Unfortunately the treatment in Ref. \cite{mottishaw:18} was not done using replicas so it 
is not obvious how it might be extended directly to the finite $N$ fully-connected $p$-spin model.

Our second argument also concerns lifting the degeneracy of the states with KRSB but this time by the  fluctuation effects about the mean-field solution. The 1RSB (and the KRSB) state do not have massless modes (null eigenvalues in their Hessian matrix) \cite{campellone:99}. For $p$-spin models with $p$ even and $p/2$ odd,(such as $p=6$) the system has time reversal invariance and one can calculate the free energy cost of creating a droplet of reversed spins with linear extent $R$  within the state \cite{moore:06b}. (The calculations of this reference need to be qualified to make it clear that they are only applicable when $p$ is even with $p/2$ odd). The droplet free energy cost varies as $ \sim \exp(-R/\xi)$, where $\xi$ is the longest length scale in the system. Thus the free energy cost  of flipping the spins in a region of size $R$ will vanish for $R \gg \xi$, implying that the state would be unstable against thermal fluctuations. However, in the presence of FRSB, one expects massless modes to occur i.e. that $\xi \to \infty$ \cite{dominicis:97}. An example of a state with massless modes is the low temperature phase of the Ising ($p=2$) spin glass,  which is a phase with FRSB  \cite{dominicis:97} and is stable against excitations  of spin-flipped droplets. We would therefore conclude that the state which exists between $T_G$ and $T^c_{\rm 1RSB}$ can only be stable against such fluctuations if it has FRSB. (This argument only holds for $p$ even with $p/2$ odd, but we suspect that for other $p$ values droplets, (but not simply time-reversed droplets) will exist to destabilize them also.)

In Sec.~\ref{2RSB} we set up the 2RSB calculations and present the main results. In Sec.~\ref{p6} we introduce  new types of $(M-p)$ models which are convenient for describing fluctuation effects when $p$ is even, the "balanced" $(M-p)$ models, and show that amongst them there is a subset, those with $p$ even and $p/2$ odd for which the calculations in \cite{moore:06b} are valid. In Sec.~\ref{discussion} we speculate what might happen in physical dimensions like $d = 3$, which are a long way away from the mean-field limit of the fully connected model studied in Sec. \ref{2RSB}. There are simulations in three dimensions which have FRSB-like features \cite{campellone:98, franz:99}. It is these studies which suggested to us that the state of 1RSB might just be a feature of the infinite $N$ fully-connected $p$-spin model. We shall also comment on the possible relevance of our work to structural glasses, (which is  the reason why one studies $p$-spin models) and speculate as to whether the fluctuations remove not only 1RSB but FRSB in three dimensions.

\section{The fully-connected $p$-spin glass model}
\label{2RSB}

In this section, we study the fully-connected $p$-spin glass model described by the Hamiltonian of Eq.~(\ref{Ham}) in the limit when the number of spins $N$ goes to infinity.
 The replicated partition function averaged over the random couplings
at inverse temperature $\beta$ can be
expressed \cite{gardner:85} as integrals over the auxiliary fields $q_{ab}$ and $\lambda_{ab}$ as
\begin{align}
 \langle Z^n\rangle =&e^{nN\beta^2/4}\int\prod_{a<b}dq_{ab}\int\prod_{a<b}d\lambda_{ab}\; \nonumber \\
 &\times \exp[-NG(q_{ab},\lambda_{ab})],
 \label{Zn}
\end{align}
where the $q_{ab}$ and $\lambda_{ab}$ are symmetric matrices with the replica indices $a,b=1,2,\cdots,n$ with
zero diagonal part and
\begin{align}
 G(q_{ab},\lambda_{ab})=&-\frac{\beta^2}{4}\sum_{a\neq b}q^p_{ab}+\frac 1 2 \sum_{a \neq b}
 \lambda_{ab}q_{ab}\nonumber \\
 &-\ln\mathrm{Tr}_{S_a}\exp[\frac 12 \sum_{a\neq b}\lambda_{ab}S_a S_b].
 \label{Znp}
\end{align}
In the large-$N$ limit, the integral in Eq.~(\ref{Zn}) is given by the saddle point values of $q_{ab}$
and $\lambda_{ab}$
The free energy per site of the $p$-spin spin glass model is then given by
\begin{equation}
 \frac{F}{N} = \frac 1\beta \lim_{n\to 0} \frac 1 n G[q_{ab},\lambda_{ab}]-\frac{\beta}4,
\end{equation}
where 
the saddle point values of $q_{ab}$ and $\lambda_{ab}$ are assumed to be used.

If we take the replica symmetric form, $q_{ab}=q$ and $\lambda_{ab}=\lambda$
for $a\neq b$, the solution to the saddle point equations are given by $\lambda=q=0$
and the free energy per site is just
\begin{equation}
 \frac{F_{\rm RS}}N= -\frac{\beta}{4}-\frac{\ln 2}{\beta}.
 \label{f_rs}
\end{equation}
The entropy per site is then given by
\begin{equation}
 \frac{S_{\rm RS}}N=\ln 2- \frac {\beta^2}{4},
\end{equation}
which becomes negative for temperature $T<T_*\equiv 1/(2\sqrt{\ln 2})\simeq 0.601$.

\subsection{One-step Replica Symmetry Breaking}

We now consider the case where the saddle point values of $q_{ab}$ and $\lambda_{ab}$ take the 1RSB form, where
$q_{ab}$ and $\lambda_{ab}$ is equal to $q$ and $\lambda$ on $n/m$ diagonal 
blocks of size $m$ and $q_0$ and $\lambda_0$
outside the blocks, respectively. 
The detailed derivations can be found elsewhere \cite{gardner:85,campellone:99} and we present the relevant results
in the following. 

In the absence of 
an external field, the 1RSB saddle points are given by $q_0=\lambda_0=0$. 
The other parameters are determined by
\begin{equation}
 \lambda=\frac {\beta^2}2 p q^{p-1},
\end{equation}
and 
\begin{equation}
 q= \frac{\int Dy\; \cosh^{m} (\sqrt{\lambda}y)\tanh^2 (\sqrt{\lambda}y)}
 {\int Dy\; \cosh^{m}(\sqrt{\lambda}y)},
 \label{q1_1rsb}
\end{equation}
where 
\begin{equation}
 \int Dz \equiv \frac 1 {\sqrt{2\pi}}\int_{-\infty}^{\infty} dz\; e^{-z^2/2}. 
\end{equation}
The free energy for the 1RSB solution is given by
\begin{align}
 \frac{F_{\rm 1RSB}}{N}=& -\frac {\beta}{4} [1+(1-m)(p-1)q^p -pq^{p-1}] \nonumber \\
& -\frac{\ln 2}{\beta} -\frac 1 {\beta m}
\ln \int Dy\; \cosh^{m}(\sqrt{\lambda} y) .
\label{f_1rsb_2}
\end{align}
There is another saddle point equation which is obtained by varying the free 
energy with respect to $m$:
\begin{align}
0&=\frac{\beta^2}4 q^p (p-1) + \frac 1 {m^2} \ln \int Dy\;
 \cosh^{m} (\sqrt{\lambda}y) \nonumber \\
&-\frac 1 {m} \frac{\int Dy\; \cosh^{m} (\sqrt{\lambda}y)
\ln(\cosh(\sqrt{\lambda}y))}{\int Dy\; \cosh^{m} (\sqrt{\lambda}y)}=0.
\label{m1_1rsb}
 \end{align}
 
We note that when $m=1$, $F_{\rm 1RSB}$ becomes equal to $F_{\rm RS}$ in Eq.~(\ref{f_rs}).
We determine the temperature $T^c_{\rm 1RSB}$ at which the two free energies are equal to 
each other by setting $m=1$ in Eqs.~(\ref{q1_1rsb}) and (\ref{m1_1rsb})
and by solving the equations for $\beta$. We have $T^c_{\rm 1RSB}\simeq 0.651$ for $p=3$. 

As noted by Gardner in Ref.~\cite{gardner:85}, the 1RSB solution is stable only when 
\begin{equation}
 \frac{q}{(p-1)\lambda}>\frac{\int Dy\; \cosh^{m-4}(\sqrt{\lambda} y)}{\int Dy\; \cosh^m(\sqrt{\lambda}y)}.
\end{equation}
If we use the 1RSB solution, this equation translates into $T>T_G$, where $T_G\simeq 0.24$ for $p=3$.

\subsection{Two-step Replica Symmetry Breaking}

We now consider the case where $q_{ab}$ takes the two-step replica symmetry breaking (2RSB) form 
with values $q_0$, $q_1$ and $q_2$ specified by the parameters
$m_1$ and $m_2$. There are $n/m_1$ diagonal blocks of size $m_1$ denoted by $B_i$, 
$i=1,\cdots,n/m_1$. Outside $B_i$, $q_{ab}=q_0$. Inside each $B_i$, $q_{ab}=q_1$ except for 
$m_1/m_2$ diagonal blocks of size $m_2$
denoted by $B_i^j$, $j=1,\cdots,m_1/m_2$, where $q_{ab}=q_2$. 
$\lambda_{ab}$ takes the same form with $\lambda_0$, $\lambda_1$ and $\lambda_2$.
The terms in the free energy are now given by
\begin{equation}
 \sum_{a\neq b}q^p_{ab}=n[(m_2-1)q^p_2 +(m_1-m_2)q^p_1 +(n-m_1)q^p_0], \label{qp2} 
 \end{equation}
 \begin{align}
 \sum_{a\neq b}\lambda_{ab}q_{ab}=&n[(m_2-1)\lambda_2 q_2 +(m_1-m_2)\lambda_1 q_1 \nonumber \\
 &+(n-m_1)\lambda_0 q_0],
 \label{lamp2}
\end{align}
and
\begin{widetext}
\begin{align}
 \frac 1 2 \sum_{a\neq b}\lambda_{ab}S_a S_b =&
 \frac {\lambda_0}2 \sum^n_{a\neq b} S_a S_b +\frac 1 2 (\lambda_1-\lambda_0)\sum_{i=1}^{n/m_1} 
 \sum_{\substack{a,b \in B_i \\ a\neq b}} S_a S_b
 +\frac 1 2(\lambda_2-\lambda_1)\sum_{i=1}^{n/m_1} \sum_{j=1}^{m_1/m_2}
 \sum_{\substack{a,b \in B^j_i \\ a\neq b}} S_a S_b ,
 \label{lss2rsb}
\end{align}
We then decouple the terms containing spins with two different replica indices 
using the Hubbard-Stratonovich transformation. After taking the trace over the spins, we obtain
the expression for the 2RSB free energy as
\begin{align}
 \frac{F_{\rm 2RSB}}N=& -\frac {\beta}{4} [1+(m_2-1)q^p_2+(m_1-m_2) q^p_1-m_1 q^p_0]
 +\frac 1 {2\beta} 
 [(m_2-1)\lambda_2 q_2 +(m_1-m_2) \lambda_1 q_1 -m_1 \lambda_0 q_0 ] \nonumber \\
 &+\frac {\lambda_2}{2\beta} 
 -\frac{\ln 2}{\beta}-\frac 1 {\beta m_1}
 \int Dz\; \ln \int Dy\; \left\{ \int Dw\; \cosh^{m_2}(\eta(z,y,w)) \right\}^{m_1/m_2},
 \label{f_2rsb_1}
\end{align}
where
\begin{equation}
 \eta(z,y,w)=\sqrt{\lambda_0}z+\sqrt{\lambda_1-\lambda_0}y+\sqrt{\lambda_2-\lambda_1}w.
\end{equation}
Varying the free energy with respect to $q_i$, we obtain 
$\lambda_i=\beta^2 p q^{p-1}_i/2$
for $i=0,1$ and 2. 

The saddle point equation obtained from varying $\lambda_0$ gives
\begin{equation}
 q_0=\int Dz\; \left[ \frac{\int Dy\; \left\{ \int Dw_1\; \cosh^{m_2}(\eta(z,y,w_1))\right\}^{\frac{m_1}{m_2}-1}
 \int Dw_2\; \cosh^{m_2}(\eta(z,y,w_2))  \tanh(\eta(z,y,w_2))}
 {\int Dy\; \left\{ \int Dw\; \cosh^{m_2}(\eta(z,y,w))\right\}^{\frac{m_1}{m_2}} } \right]^2 .
\end{equation}
We can show that $q_0=0$ and $\lambda_0=0$ are solutions to the above equation and set them to zero from now on. 
Varying the free energy with respect to $\lambda_1$ and $\lambda_2$ gives, respectively,
\begin{equation}
 q_1= \frac{\int Dy\; \left\{ \int Dw_1\; \cosh^{m_2}(\zeta(y,w_1))\right\}^{\frac{m_1}{m_2}-2}
 \left\{ \int Dw_2\; \cosh^{m_2}(\zeta(y,w_2))  \tanh(\zeta(y,w_2)) \right\}^2 }
 {\int Dy\; \left\{ \int Dw\; \cosh^{m_2}(\zeta(y,w))\right\}^{\frac{m_1}{m_2}} } 
 \label{q1_2rsb}
 \end{equation}
 and
 \begin{equation}
 q_2= \frac{\int Dy\; \left\{ \int Dw_1\; \cosh^{m_2}(\zeta(y,w_1))\right\}^{\frac{m_1}{m_2}-1}
 \int Dw_2\; \cosh^{m_2}(\zeta(y,w_2))  \tanh^2(\zeta(y,w_2))}
 {\int Dy\; \left\{ \int Dw\; \cosh^{m_2}(\zeta(y,w))\right\}^{\frac{m_1}{m_2}} }  ,
 \label{q2_2rsb}
 \end{equation}
 where
 \begin{equation}
 \zeta(y,w)=\sqrt{\lambda_1}y+\sqrt{\lambda_2-\lambda_1}w.
 \label{zeta}
 \end{equation}
 Using $q_0=\lambda_0=0$, we can rewrite the free energy, Eq.~(\ref{f_2rsb_1}) as 
\begin{align}
\frac{F_{\rm 2RSB}}{N}   
 =& -\frac {\beta}{4} \Big[1+(p-1)\{ (1-m_2)q^p_2+(m_2-m_1)q^p_1\}  -pq^{p-1}_2\Big]  \nonumber \\
& -\frac{\ln 2}{\beta} -\frac 1 {\beta m_1}
\ln \int Dy\; \left\{ \int Dw\; \cosh^{m_2}(\zeta(y,w)) \right\}^{m_1/m_2} . \label{f_2rsb_2}
\end{align} 
\end{widetext}
 
We evaluate the 2RSB free energy by solving the above saddle point equations numerically. 
There are two more saddle point equations which can be obtained by 
varying the free energy with respect to $m_1$ and $m_2$. We find, however, that finding numerical solution
of the full set of these coupled saddle point equations is rather tricky. 
It is more convenient to 
solve the two equations Eqs.~(\ref{q1_2rsb}) and (\ref{q2_2rsb}) first for given $m_1$ and $m_2$ at
fixed inverse temperature $\beta$. We then evaluate $F_{\rm 2RSB}$ from Eq.~(\ref{f_2rsb_2}) and find the
values of $m_1$ and $m_2$ that maximize the free energy.
Since $1\le m_2 \le m_1 \le n$, we have to have $0\le m_1 \le m_2 \le 1$ in 
the limit $n\to 0$. 

Before presenting the 2RSB solution, we first look at the trivial limits of these equations. 
We first note that $q_1=0$ is always a solution to Eq.~(\ref{q1_2rsb}), since 
the integrals become independent of $y$ with $\zeta=\sqrt{\lambda_2}w$. 
When $q_1=\lambda_1=0$, Eq.~(\ref{q2_2rsb}) for $q_2$
becomes exactly the same as Eq.~(\ref{q1_1rsb}) for the 1RSB solution if we regard $q_2$ and $m_2$
as the 1RSB parameters $q$ and $m$, respectively.
The 2RSB free energy, Eq.~(\ref{f_2rsb_2}) also reduces to the one for the 1RSB scheme, Eq.~(\ref{f_1rsb_2}).
We note that this solution is independent of $m_1$. 

There is another limit where the equations reduce to the 1RSB ones. It is the case where $q_1=q_2$.
We can indeed check this is a solution if we take $\lambda_1=\lambda_2$
in Eqs.~(\ref{q1_2rsb}) and (\ref{q2_2rsb}). We find 
the integrals over $w$
can be done easily as the integrands 
become independent of $w$. The right hand sides of the two equations become identical
to each other and also to Eq.~(\ref{q1_1rsb}) for the 1RSB case if we regard $q_1=q_2$ and
$m_1$ as the 1RSB parameters $q$ and $m$, respectively. We can also check in this limit that 
the free energy reduces to the 1RSB one. In this case, the solution is independent of $m_2$. 

\begin{figure}
\includegraphics[width=0.55\textwidth]{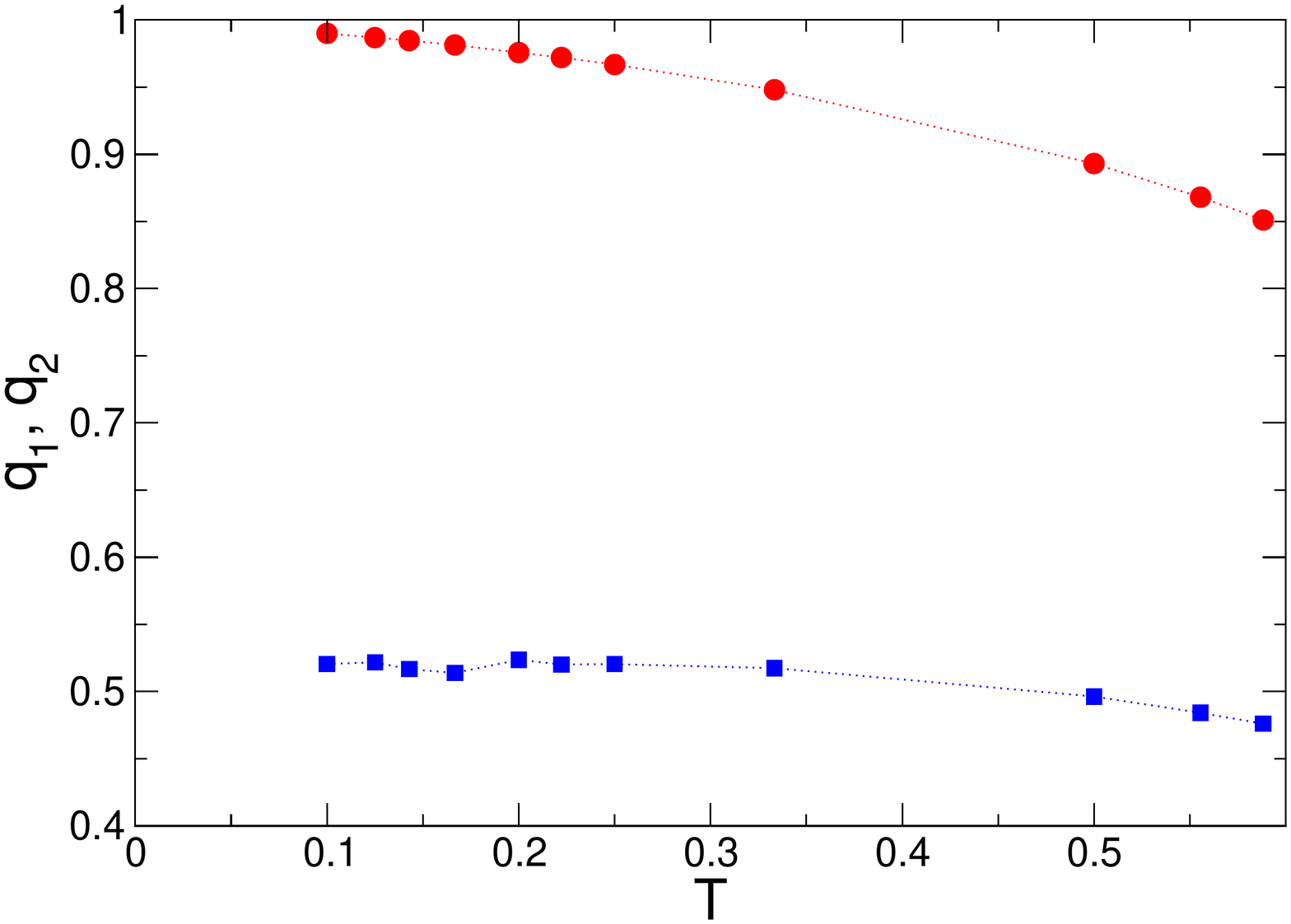} 
\caption{Type-I 2RSB solutions, $q_1$ and $q_2$ as functions of temperature $T$ for $p=3$. 
Squares (blue) and circles (red) correspond
to $q_1$ and $q_2$, respectively.}
\label{fig:type1_q}
\end{figure}

The 2RSB solution we look for in this paper therefore corresponds to the case 
where $0<q_1<q_2$. At fixed temperature, we solve Eqs.~(\ref{q1_2rsb}) and (\ref{q2_2rsb}) numerically
for given values of 
$m_1$ and $m_2$. We find that there are two different types of  2RSB solutions, which we denote
by type-I and type-II in the following. The type-I solution is characterized by $q_1$ 
being much smaller than $q_2$, and
it exists in the entire temperature range below $T^c_{\rm 1RSB}$. At fixed temperature, we calculate the free energy
for the type-I solution 
for various values of $m_1$ and $m_2$. For given $m_1$, we find that the free energy is always
a decreasing function of $m_2$($\ge m_1$) so that the maximum free energy 
for given $m_1$ is always at $m_2=m_1$. As we change the value of $m_1$, 
we find that the maximum occurs when $m_1$ is equal to the 1RSB solution $m(T)$ at that temperature.
In Fig.~\ref{fig:type1_q}, we plot the type-I 2RSB solutions, $q_1$ and $q_2$ as a function of temperature. 
In the zero-temperature limit, $q_2$ approaches 1, while $q_1$ saturates to a constant value around 0.52.

We can see that the type-I 2RSB solution with $m_1=m_2$ actually reproduces the 1RSB free energy in 
the following way. 
If we put $m_1=m_2$ in Eq.~(\ref{q2_2rsb}), it becomes
\begin{equation}
 q_2= \frac{\int Dy\; 
 \int Dw\; \cosh^{m_2}(\zeta(y,w))  \tanh^2(\zeta(y,w))}
 {\int Dy\; \int Dw\; \cosh^{m_2}(\zeta(y,w)) }
 \label{q2m12}
\end{equation}
If we use the notation 
\begin{equation}
\sigma_1 =\sqrt{\lambda_1},~~~~~\sigma_2 =\sqrt{\lambda_2-\lambda_1},
\end{equation}
and the vector notation $\vec{\sigma}=(\sigma_1,\sigma_2)$ 
and $\vec{r}=(y,w)$ for the integration variable $y$ and $w$, then
$\zeta(y,w)=\vec{\sigma}\cdot\vec{r}$ and we can rewrite
Eq.~(\ref{q2m12}) as
\begin{equation}
 q_2 =\frac{\int D^2\vec{r}\; \cosh^{m_2}
 (\vec{\sigma}\cdot\vec{r}) \tanh^2 (\vec{\sigma}\cdot \vec{r})}
 {\int D^2\vec{r}\; \cosh^{m_2}(\vec{\sigma}\cdot\vec{r}) },
 \label{q2_2rsb_3}
\end{equation}
where 
\begin{equation}
\int D^2\vec{r} \equiv \int Dy \int Dw =\int  \frac{d^2\vec{r}}{2\pi} \;e^{-\vert\vec{r}\vert^2/2} .
\end{equation} 
Since the right hand side of Eq.~(\ref{q2_2rsb_3}) depends only on 
$\vert\vec{\sigma}\vert$, it determines the value of $\vert\vec{\sigma}\vert$
or 
\begin{equation}
 q_2=\left(\frac{2\lambda_2}{p\beta^2}\right)^{1/(p-1)}=
 \left(\frac{2\vert\vec{\sigma}\vert^2}{p\beta^2}\right)^{1/(p-1)}
\end{equation}
for given $m_1=m_2$ and $\beta$. We note that this equation is exactly the same as the 1RSB
one Eq.~(\ref{q1_1rsb}) with $q_2$ and 
$\vert\vec{\sigma}\vert$ playing the role of $q$ and $\sqrt{\lambda}$, respectively, for the 1RSB case.
The 2RSB free energy, Eq.~(\ref{f_2rsb_2}) also becomes the 1RSB one, Eq.~(\ref{f_1rsb_2}) when $m_1=m_2$.
Therefore $q_2$ shown in Fig.~\ref{fig:type1_q} is the same as $q$ for the 1RSB scheme. 
On the other hand, the equation for $q_1$, Eq.~(\ref{q1_2rsb}) becomes decoupled from the rest
of the equations for $m_1=m_2$, and in particular does not appear in the free energy expression. 
The solution of 
this equation for $q_1$ is shown in Fig.~\ref{fig:type1_q}.

\begin{figure}
\includegraphics[width=0.55\textwidth]{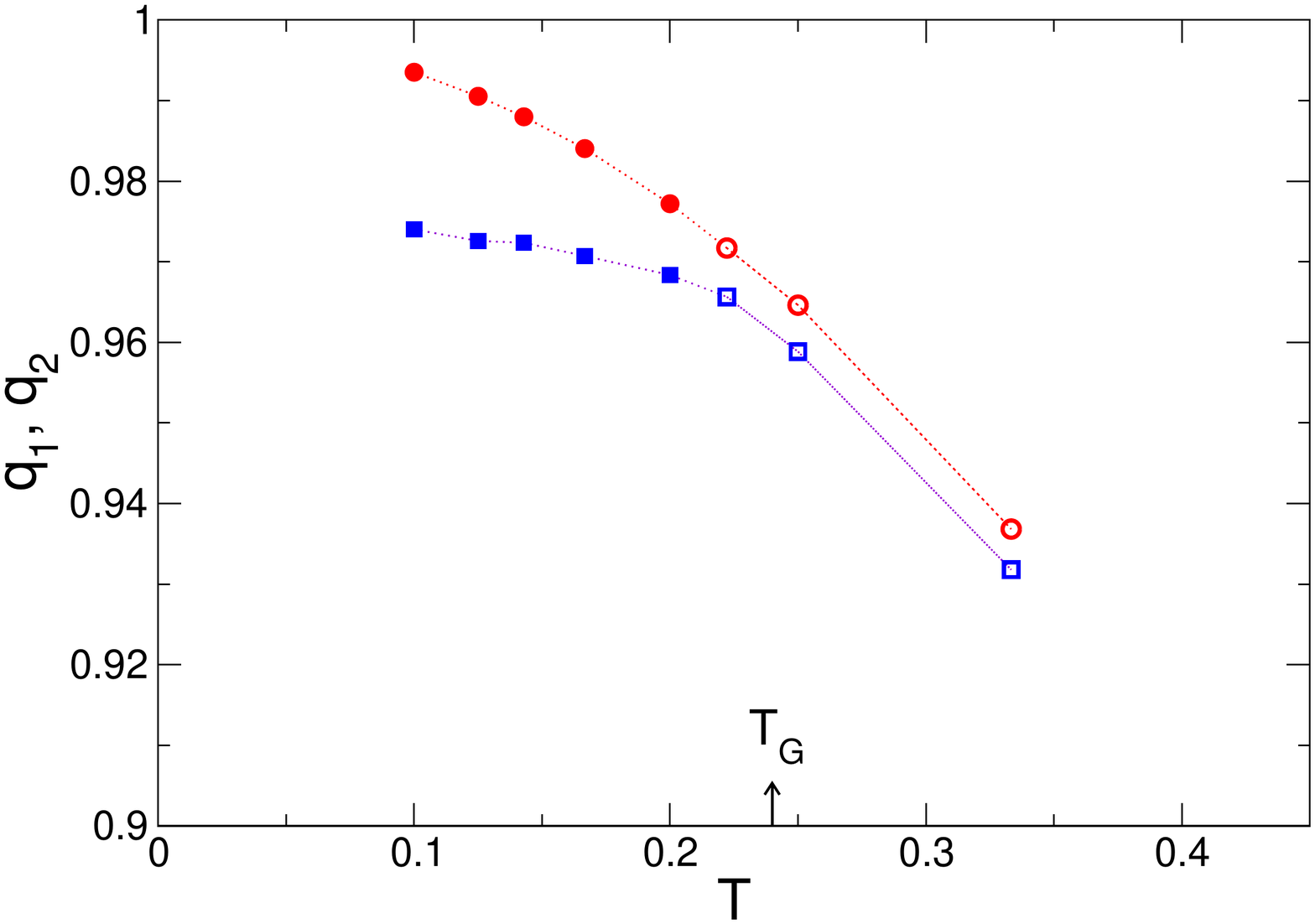} 
\caption{Type-II 2RSB solutions, $q_1$ and $q_2$ as functions of temperature $T$ for $p=3$. 
Squares (blue) and circles (red) correspond
to $q_1$ and $q_2$, respectively. The filled and open symbols correspond to the cases $m(T)<m^*_1(T)$
and $m(T)>m^*_1(T)$, (see Fig.~\ref{fig:type12_m}).  $T_G$ is the Gardner temperature}
\label{fig:type2_q}
\end{figure}

\begin{figure}
\includegraphics[width=0.55\textwidth]{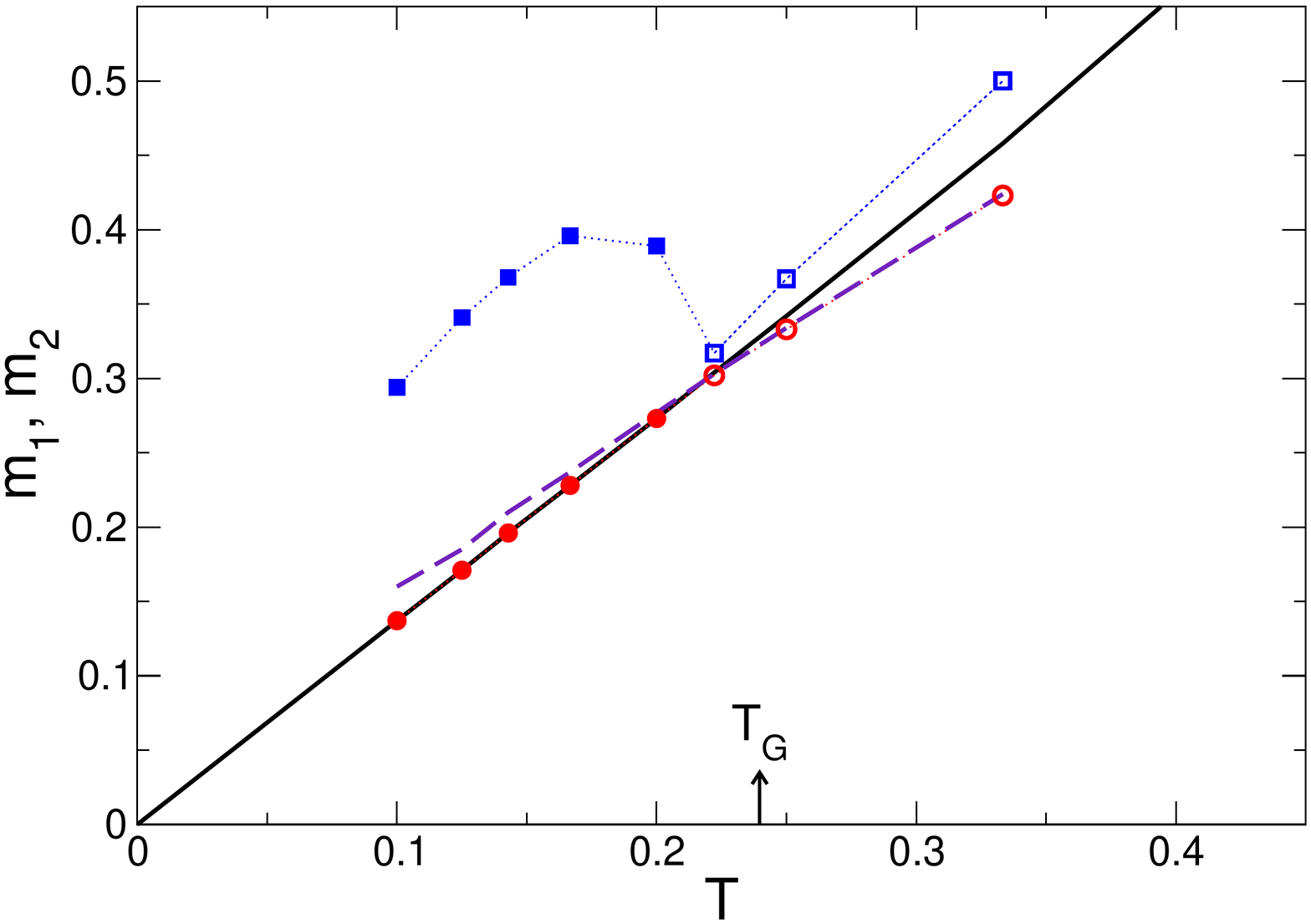} 
\caption{Type-II 2RSB solutions, $m_1$ and $m_2$ as functions of temperature $T$ for $p=3$. 
Circles (red) and
and squares (blue) correspond
to $m_1$ and $m_2$, respectively. The solid line is the 1RSB solution $m(T)$.
The dashed line is $m_1^*(T)$ above which the type-II 2RSB solution disappears.
The filled and open symbols correspond to the cases $m(T)<m^*_1(T)$
and $m(T)>m^*_1(T)$, respectively. $T_G$ is the Gardner temperature.}
\label{fig:type12_m}
\end{figure}

\begin{figure}
\includegraphics[width=0.55\textwidth]{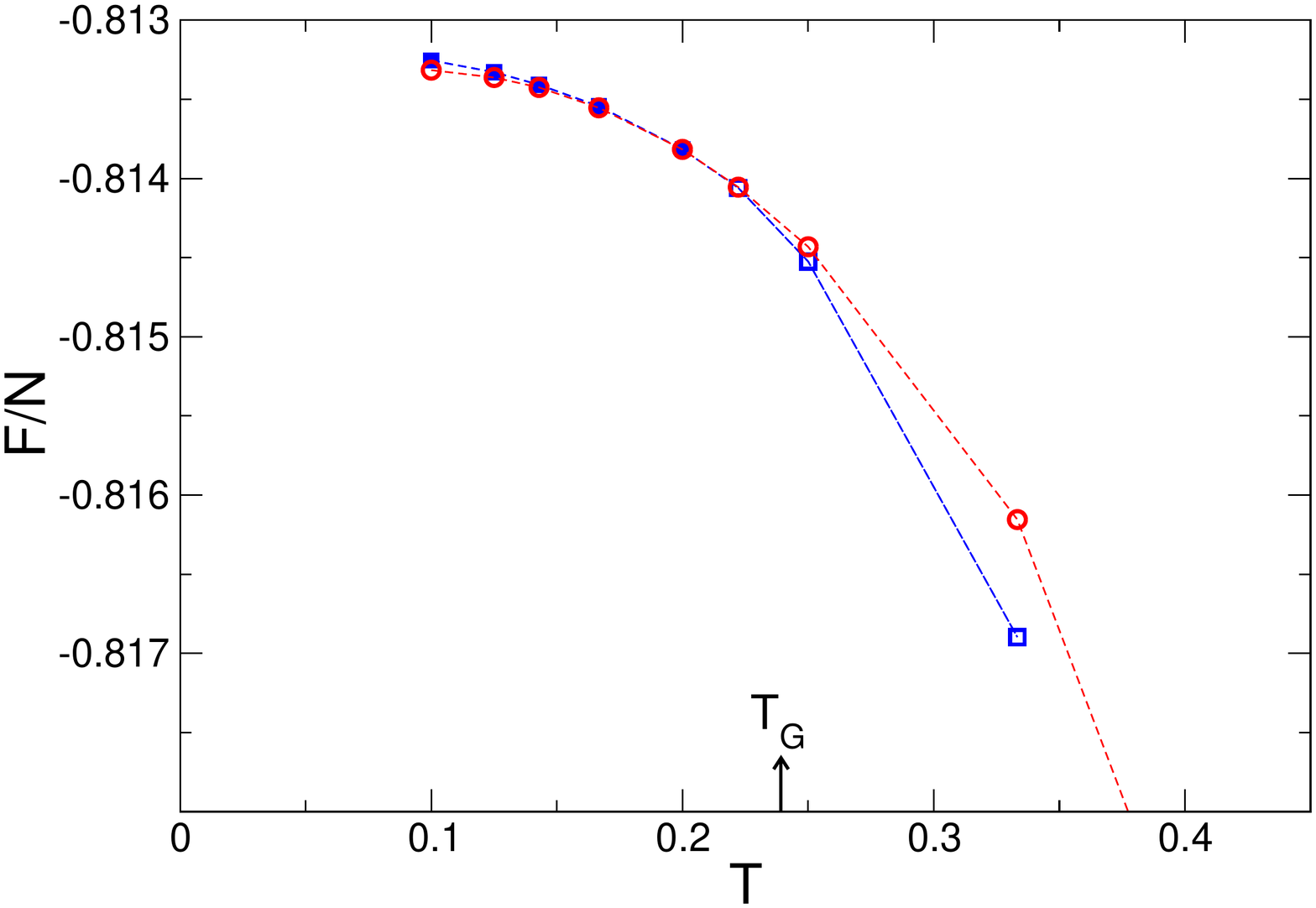} 
\caption{The free energy per site of the type-II 2RSB solution (squares)
vs. the 1RSB free energy per site (circles). }
\label{fig:type12_free}
\end{figure}

The type-II 2RSB solution is characterized by $q_1$ being closer to $q_2$ than the type-I counterpart
as shown in Fig.~\ref{fig:type2_q}.
The type-II solution is certainly the relevant solution for  temperatures below $T_G$. 
At fixed temperature,
we again calculate the type-II 2RSB free energy for various values of $m_1$ and $m_2$. 
For given $m_1$, the maximum free energy occurs at some value of $m_2$ which is 
larger than $m_1$. If we now change $m_1$, we find that the type-II solution ceases to exist
for $m_1$ larger than some value $m^*_1(T)$. At low temperatures, 
this limiting value $m^*_1(T)$ is larger than the 1RSB values $m(T)$. 
We find that the free energy achieves a
maximum when $m_1$ is equal to the 1RSB value $m(T)$ and at some $m_2$ which is larger than $m_1=m(T)$.
These values of $m_1$ and $m_2$ are shown in Fig.~\ref{fig:type12_m} as filled symbols. 
At relatively high temperatures, $m^*_1(T)$ is smaller than $m(T)$. That is, before 
$m_1$ reaches the 1RSB value $m(T)$, the solution disappears. Then the maximum free energy 
just corresponds
to $m_1=m^*_1(T)$ and $m_2$ being equal to some value larger than $m_1=m^*_1(T)$. These are shown in 
Fig.~\ref{fig:type12_m} as open symbols. As we can see from Fig.~\ref{fig:type2_q} and  Fig.~\ref{fig:type12_m}, the temperature dependence 
of the parameters is quite different for temperature above and below $T_G$. 
The change of behavior actually occurs at a temperature $T\simeq 0.22$ for $p=3$,
which is quite close to but just below $T_G$. (We suspect that for KRSB as $K$ grows, the change of behavior will approach $T_G$.)   
This alteration in behavior is also reflected in the type-II free energy and is shown in 
Fig.~\ref{fig:type12_free}. 
The 2RSB free energy is larger than the 1RSB one for $T< 0.22 \approx T_G$, but becomes smaller
than the 1RSB value at temperatures $T >T_G$. It is clear that the type-II solution is just the 2RSB approximation to the state with FRSB which is expected below $T_G$ \cite{gardner:85}. What is perhaps surprising is that this solution has an extension into the temperature region $T > T_G$. Our numerical work fails to find it above a  temperature $T^{**}$.  Note that $q_1 \to q_2$ as $T$ grows towards $T_{\rm 1RSB}^c$. If $q_1=q_2$ the $P(q)$ of 2RSB in  Eq.~(\ref{Pq2RSB}) reduces to the 1RSB form of Eq.~(\ref{Pq1RSB}). The iterative procedure used locks onto this 1RSB solution above $T^{**}$ rather than the 2RSB solution. 

We have restricted our studies of the 2RSB state to the physical region where $m_2 \ge m_1$ so that the probability associated with the delta function in Eq.~(\ref{Pq2RSB}) at $q =q_1$ remains positive. There are solutions in the unphysical region where $m_2 < m_1$ but they cannot be of physical significance. However, in a different but related context, solutions with negative probabilities have been discussed in the literature  \cite{kurchan:93,Baviera:15} and related to the properties of metastable states or barriers. The type-I solution is stationary (a maximum) along the line $m_1=m_2$, but not a true saddle point in the full physical region of the $m_1,m_2$ space. The type-II solution is a genuine stationary solution in these variables provided $T < T_G$. The ideal solution would be a saddle point rather than one having to be constrained \lq\lq by hand" to lie in the physical region. However, we expect that this difficulty will go away in the $K \to \infty$ limit when the system is stabilized by finite size effects or by fluctuations. It is invariably the case that for a system which has FRSB is approximated by KRSB there are unsatisfactory features to be found in that level of approximation.

\section{The case of $p$ even, $p/2$ odd}
\label{p6}
The results in Sec.~\ref{2RSB} are for the fully-connected $p$-spin model which provides the mean-field limit for the $p$-spin model. To go beyond mean-field theory many workers have found it convenient to study a generalization of the $p$-spin model, the so-called $(M-p)$ models in which one has $M$ different types of Ising spins coupled by short-range (usually nearest-neighbor) $p$-spin interactions \cite{campellone:99,drossel:02,moore:06,caltagirone:11,campellone:98,yeo:12}. In this section we introduce a model of this kind which provides a convenient way of discussing fluctuation effects when $p$ is even. We refer to it as the  "balanced" $(M-p)$ model as in this model if $x$ and $y$ denote two nearest-neighbor points on a lattice there are always $p/2$ spins on site $x$ and $p/2$ spins on site $y$. In the usual $(M-p)$ model, the number of spins on either $x$ or $y$ can range from $1$ up to $p-1$. The big advantage of the balanced model is that with it there is no need to introduce  hard and soft modes, as in \cite{drossel:02,caltagirone:11}.

Denote the $M$ different spins which can be placed at site $x$ by $S_i(x)$, $i=1,2,\cdots,M$. We shall focus on the case when $p=6$.
Let us define for this case
\begin{widetext}
\begin{align}
R_{\alpha\beta} (x)& \equiv \frac 1 {M^3} \sum_{i_1<i_2<i_3}^M  S^\alpha_{i_1}(x)S^\beta_{i_1}(x)
S^\alpha_{i_2}(x)S^\beta_{i_2} (x) S^\alpha_{i_3}(x) S^\beta_{i_3} (x)  \label{eq1} \\
&=  \frac{1}{3! M^3}  \sum_{\substack{i_1,i_2,i_3 \\ \mathrm{distinct}}}^M  S^\alpha_{i_1}(x)S^\beta_{i_1}(x)
S^\alpha_{i_2}(x)S^\beta_{i_2} (x) S^\alpha_{i_3}(x)S^\beta_{i_3}(x)  \nonumber \\
&=\frac 1 6 \left(\frac 1 M \sum_i^M  S^\alpha_i (x) S^\beta_i (x)  \right)^3 -\frac{3M-2}{6M^3}\sum_i^M S^\alpha_i (x) S^\beta_i (x)   \nonumber \\
&\simeq \frac 1 6 Q^3_{\alpha\beta}(x) ,
\end{align}
where in the large-$M$ limit, the second term, which comes from the diagonal elements, is subleading
and we have defined
\begin{equation}
Q_{\alpha\beta} (x) \equiv \frac 1 M \sum_i ^M  S^\alpha_i (x) S^\beta_i  (x). \label{order}
\end{equation}

The Hamiltonian for the balanced $(M-p)$ model for $p=6$ is

\begin{equation}
H=-\frac 1 2 \sum_{x\neq y} \sum_{i_1<i_2<i_3}^M \sum_{j_1<j_2<j_3}^M J^{(\bm{i},\bm{j})}_{(x,y)} \;S_{i_1}(x)
S_{i_2}(x) S_{i_3}(x) S_{j_1}(y)
S_{j_2}(y) S_{j_3}(y) ,
\end{equation}
and each $J^{(\bm{i},\bm{j})}_{(x,y)}$ is chosen independently from a Gaussian distribution of variance $J^2/M^{p-1}$ and it only couples sites $x$ and $y$ which are nearest-neighbors.
The replicated partition function for the even $p$ balanced model basically looks like

\begin{align}
\overline{Z^n}&\sim \mathrm{Tr} \exp\left[ \frac 1 2 \sum_{x,y} K(x,y) \sum_{\alpha<\beta} R_{\alpha\beta}(x) R_{\alpha\beta}(y)\right] \nonumber \\
&\sim \int \prod_{\substack{x \\ \alpha <\beta}} d\phi_{\alpha\beta}(x)\; 
\exp\left [-\frac 1 2 \sum_{x,y} \sum_{\alpha<\beta} \phi_{\alpha\beta}(x)K^{-1}(x,y) \phi_{\alpha\beta}(y)\right] \nonumber \\
&~~~~~~~~~~\times\mathrm{Tr} \exp\left[\sum_x\sum_{\alpha<\beta} \phi_{\alpha\beta}(x)R_{\alpha\beta}(x)\right],
\end{align}
where $K(x,y)$ is zero if $x$ and $y$ are not nearest-neighbors.   Reference \cite{caltagirone:11} distinguishes the case of $p$ even and $p/2$ even, (such as $p=4$) and the situation when $p$ is even and $p/2$ is odd, (such as $p=6$). In the case of $p=4$ the field theory in $\phi_{\alpha \beta}$ has terms like $w_2 \phi_{\alpha  \beta}^3$ with $w_2$ non-zero, which can be obtained by tracing out $R_{\alpha \beta}$. For the case $p=6$ coefficients like $w_2$ are all zero. In other words, the field theory in terms of $\phi_{\alpha \beta}$ for $p=6$ is manifestly time reversal invariant and rather like that of the zero field Ising spin glass model. For $p=4$ the field theory is more like that of the Ising spin glass model in a field.

\subsection{The fully connected limit of the balanced model}
For the fully connected $p=6$ balanced model 
 each $J^{(\bm{i},\bm{j})}_{(x,y)}$ is chosen independently from the Gaussian distribution with zero mean
and the variance
\[
\frac {J^2}{NM^{p-1}} =\frac {J^2}{NM^5}
\]
for $p=6$. We shall now show that when $M$ is large the fully-connected balanced model reduces to the $p=6$ model in Eq.~(\ref{Znp}).

The replicated and bond-averaged partition function is 
\begin{equation}
\overline{Z^n}=e^{nNC} \mathrm{Tr} \exp\left[ \frac {\beta^2 J^2 M}{4N} \sum_{x\neq y}  \sum_{\alpha\neq \beta} R_{\alpha\beta}(x) R_{\alpha\beta}(y)\right] ,
\end{equation}
where $C$ is a constant coming from the diagonal replica terms and $R_{\alpha\beta}(x)$ is a shorthand notation 
for the collection of spins given in Eq.~(\ref{eq1}).
This can be rewritten using the integral representation (via $\Lambda_{\alpha \beta}$) of the delta function constraint enforcing
$R_{\alpha\beta}=\frac 1 {N}\sum_{x} R_{\alpha\beta}(x)$ or

\begin{equation}
NM R_{\alpha\beta}=\frac 1{M^{2}}\sum_{x}
\sum_{i_1<i_2<i_3}^M  S^\alpha_{i_1}(x)S^\beta_{i_1}(x)
S^\alpha_{i_2}(x)S^\beta_{i_2} (x) S^\alpha_{i_3}(x) S^\beta_{i_3} (x)
\end{equation}

as
\begin{align}
\overline{Z^n}=e^{nNC}&\int\prod_{\alpha<\beta}dR_{\alpha\beta} \int\prod_{\alpha<\beta}d\Lambda_{\alpha\beta}\;
\exp\Big[ \frac{\beta^2 J^2 NM}{4} \sum_{\alpha\neq \beta} R^2_{\alpha\beta} \\
&- NM \sum_{\alpha<\beta}\Lambda_{\alpha\beta}R_{\alpha\beta} +N\ln \mathrm{Tr}_{S} \;e^{A(\Lambda)}\Big],
\end{align}

where 
\begin{equation}
A(\Lambda)=\sum_{\alpha<\beta} \frac{\Lambda_{\alpha\beta}}{M^2}  
\sum_{i_1<i_2<i_3}^M  S^\alpha_{i_1}S^\beta_{i_1}
S^\alpha_{i_2} S^\beta_{i_2}  S^\alpha_{i_3} S^\beta_{i_3} \label{action}
\end{equation}
and we have neglected the diagonal terms in sites which are subleading in the large-$N$ limit. 
In the large-$N$ limit, the integral is dominated by the saddle point values. 
Varying with respect to $R_{\alpha\beta}$, we have 
\begin{equation}
\Lambda_{\alpha\beta}=\beta^2 J^2  R_{\alpha\beta}.
\end{equation}
The replicated partition function can now be written as
\begin{equation}
\overline{Z^n}=e^{nNC}\int\prod_{\alpha<\beta}dR_{\alpha\beta} \; 
\exp\big[-\frac{\beta^2 J^2 NM}{2} \sum_{\alpha <\beta}R^2_{\alpha\beta} + N\ln \mathrm{Tr}_{S} \;e^{A(R)} \big],
\end{equation}
From Eq.~(\ref{action}), we have
\begin{align}
A(R) &= \beta^2 J^2 M \sum_{\alpha<\beta} \frac{R_{\alpha\beta}}{M^3}  
\sum_{i_1<i_2<i_3}^M  S^\alpha_{i_1}S^\beta_{i_1}
S^\alpha_{i_2} S^\beta_{i_2}  S^\alpha_{i_3} S^\beta_{i_3} \\
&=  \beta^2 J^2 M\sum_{\alpha<\beta} \frac{R_{\alpha\beta}}{3! M^3}  \sum_{\substack{i_1,i_2,i_3 \\ \mathrm{distinct}}}^M  S^\alpha_{i_1}S^\beta_{i_1}
S^\alpha_{i_2}S^\beta_{i_2}  S^\alpha_{i_3}S^\beta_{i_3} \nonumber \\
&\simeq \frac{\beta^2 J^2 M}{6} \sum_{\alpha<\beta}R_{\alpha\beta}
\left(\frac 1 M \sum_i^M  S^\alpha_i S^\beta_i   \right)^3 ,
  \nonumber 
\end{align}
where we have neglected the subleading terms in the large-$M$ limit. 
We again insert the delta function constraint using $\lambda_{\alpha \beta}$ to enforce
\begin{equation}
q_{\alpha\beta}=\frac 1 M \sum_i^M  S^\alpha_i S^\beta_i
\end{equation}
then we have
\begin{align}
\overline{Z^n}=e^{nNC}&\int\prod_{\alpha<\beta}dR_{\alpha\beta} 
\int\prod_{\alpha<\beta}dq_{\alpha\beta} \int\prod_{\alpha<\beta}d\lambda_{\alpha\beta}\; 
\exp\Big[-\frac{\beta^2 J^2 N M}{2} \sum_{\alpha <\beta}R^2_{\alpha\beta} \nonumber \\
&+ \frac{\beta^2 J^2 NM}{6} \sum_{\alpha<\beta}R_{\alpha\beta}q^3_{\alpha\beta}
-NM\sum_{\alpha<\beta}\lambda_{\alpha\beta}q_{\alpha\beta}
+NM\ln\mathrm{Tr}\;e^{\tilde{A}(\lambda)}\Big],
\end{align}
\end{widetext}
where
\begin{equation}
\tilde{A}(\lambda)=\sum_{\alpha<\beta}\lambda_{\alpha\beta}S^\alpha S^\beta
\end{equation}
The saddle point equation for $R_{\alpha\beta}$ gives
\begin{equation}
R_{\alpha\beta}=\frac 1 6 q^3_{\alpha\beta}
\end{equation}
and we have
\begin{equation}
\overline{Z^n}=e^{nNC} 
\int\prod_{\alpha<\beta}dq_{\alpha\beta} \int\prod_{\alpha<\beta}d\lambda_{\alpha\beta}\; 
\exp\Big[NMG(q,\lambda)\Big],
\end{equation}
where
\begin{equation}
G(q,\lambda)=\frac{\beta^2 J^2 }{72} \sum_{\alpha<\beta}q^6_{\alpha\beta}
-\sum_{\alpha<\beta}\lambda_{\alpha\beta}q_{\alpha\beta}
+\ln\mathrm{Tr}\;e^{\tilde{A}(\lambda)}.
\end{equation}
This is the standard expression for the fully connected 6-spin model 
if we change $J^2\to 36 J^2$; compare Eq.~(\ref{Znp}). (The first term 
is $(\beta^2 J^2 /2)\sum_{\alpha<\beta}q^p_{\alpha\beta}$ for the $p$-spin model.)
The saddle point equations
are
\begin{align}
&\lambda_{\alpha\beta}=\frac{\beta^2 J^2}{12} q^5_{\alpha\beta} \\
&q_{\alpha\beta}=\langle S^\alpha S^\beta\rangle_{\tilde{A}}.
\end{align}

\subsection{Instability of the 1RSB state against thermal fluctuations of time reversed droplets for $p$ even, $p/2$ odd}

In Ref.~\cite{moore:06b} one of us argued that the 1RSB state in the $M-p$ model is unstable against thermal fluctuations, and so cannot exist outside the mean-field limit. The argument was based on calculating the interface free energy obtained by reversing the signs of $J_{(x,y)}^{(\bm{i},\bm{j})}$ for the couplings across a $d-1$ dimensional hyperplane. A system like the balanced $(M-p)$ model with $p=6$ has the features for which the calculation of Ref.~\cite{moore:06b} is valid. For odd values of $p$ and for $p$ even with $p/2$ even (such as $p=4$) (i.e. for situations where $w_2 \ne 0$) the arguments of this reference need modifying. We suspect that there are fluctuations which will also destabilize their 1RSB state, but it is only for the case of $p$ even, $p/2$ odd such as $p=6$ that we can explicitly identify the destabilizing fluctuations as time-reversed droplets and do the calculations which show that in the 1RSB state these droplets cost very little energy so that thermal excitation of them will destroy the 1RSB order.

The key observation is that for $p=6$ the field theory has $w_2=0$ i.e.\ it has the same features as the Ising spin glass in zero applied field. This is not the case for $p=4$ which has $w_2$ non-zero. For it flipping the sign of the bonds across a $d-1$ dimensional hyperplane induces complicated changes to the state whereas for $p=6$ the field $\phi_{\alpha \beta}$ merely has a sign change, just as for the Ising case studied in \cite{timo:03} which were carried over into \cite{moore:06b}. We will not repeat the details of the calculation here. The essence of the calculation is that the interface free energy will be vanishingly small in any state which is \textit{not marginal}. In the 1RSB state (or KRSB state with K finite) there are no massless modes. We deduce from this that it is only in the mean-field limit that the 1RSB state can be stable for $p=6$. In any finite dimension it will be unstable. In its place we believe that there is a discontinuous transition to a state with FRSB. In states with FRSB there are marginal modes which stabilize the system against destruction by time reversed thermal fluctuations.

\section{discussion}
\label{discussion}

We know of no simulations of $(M-p)$ models with $p=6$. There are, however, simulations of models of $(M-p)$ models with $p=4$  in Ref.~\cite{campellone:98}. When $p=4$ there is time-reversal invariance just as for $p=6$, so its $P(q)$ in the 1RSB state will have support over the full range $-1 < q <1$. For models with time-reversal invariance Eq.~(\ref{Pq1RSB}) $P(q)$  at mean-field level becomes (with $q_0=0$)
\begin{equation}
P(q)=m \delta(q)+(1-m)/2 \delta(q-q_1)+(1-m)/2 \delta(q+q_1)
\end{equation}
Note that here $q$ relates to the overlap of two copies of the system as in Eq.~(\ref{twocopyq}). The delta functions in the simulations of Ref. \cite{campellone:98} were smeared out, although it is not clear whether this was due to finite size effects in the system or arising because the system is just not in the 1RSB state.

There are other simulational studies on the $p=4$ model in three dimensions which reveal striking differences with the results which might have been expected based on studies of the fully-connected $p$-spin model. Franz and Parisi \cite{franz:99} found for both $M=3$ and $M=4$ that as the temperature is reduced in the high temperature paramagnetic phase there was a growing (possibly diverging) length scale and susceptibility. In both the RS and 1RSB phases the length scales can be calculated  and at least at the level of Gaussian fluctuations about the mean-field solution  no  divergence as $T \to T_{\rm 1RSB}^c$ \cite{campellone:99} arises. (The study of the length scales in the $p$-spin type of model at Gaussian level is not straightforward \cite{Nieuwenhuizen:96, Ferrero:96, campellone:99}, because one has to consider fluctuations in the size of the blocks in the 1RSB matrix. It may also be further complicated by the need to allow for the degeneracy of the free energy of states with KRSB as found in this paper.) At Gaussian level the length scales are determined by the eigenvalues of the Hessian matrix: For the RS and 1RSB states there are no null eigenvalues and therefore no long length scales \cite{campellone:99}.

 Campellone et al.~\cite{campellone:98} suggested  that the difference between their results and the mean-field picture was due to non-perturbative contributions which had produced in $P(q)$ features usually associated with that of full replica symmetry breaking (FRSB), but did not provide any details of how this might arise. We have  argued that fluctuations around the mean-field solution will convert the order below $T^c_{\rm 1RSB}$ from 1RSB to FRSB in finite dimension, but we too are unable to provide a quantitative treatment.

If the discontinuous transition at $ T_{\rm 1RSB}^c$ is to a state which has FRSB, a number of questions arise. One concerns the existence of the Gardner transition \cite{gardner:85,gross:85} which is a continuous transition from within the 1RSB state to a state with FRSB. As we are claiming that there is already FRSB present for $T < T_{\rm 1RSB}^c$, it becomes a moot point as to whether in such a situation there will still be a Gardner transition. 

The field theory of the balanced $(M-p)$ models is in the variable $\phi_{ab}$. This field theory has quite different physics for the cases $p=4$ and $p=6$ where in the latter case $w_2=0$. The Gardner transition is to a state similar to that expected from the Ising model in a field \cite{gardner:85}. It is difficult to see how that would happen in finite dimensions for the $p=6$ balanced model.

Another question concerns suspicions that FRSB may not actually  exist in dimensions $d < 6$ \cite{wang:18,wang:17,moore:11}. That might be the reason why within the Migdal-Kadanoff RG procedure applied to the $M-p$ spin model with $p=3$ and $M=2$ and $M=3$ in dimensions $d = 3$ and $d =4$  we failed to see \textit{any} phase transition \cite{yeo:12,drossel:02}  as the temperature was reduced. There was, however, evidence of an ``avoided'' transition as the correlation length grew considerably, just as was reported in the direct simulations of \cite{campellone:98,franz:99}. An  avoided
transition is  one at which  the correlation  length grows but not  to
infinity as the temperature is reduced.  The $p$-spin type of model seems in low dimensions i.e. $d=3$ or $d=4$, to have similarities with the Ising spin glass in an external field. Thus the effect of fluctuations around the mean-field solution might be to remove the discontinuous transition completely in low dimensions,  though that might occur for $d > 6$ where we expect FRSB to survive.  

Since $p$-spin models are only of interest by virtue of their possible relevance to structural glasses and  RFOT theory, so we shall comment on possible implications of this avoided transition for them. The length scale in the high-temperature phase in structural glasses  is that of the point-to-set length scale $\xi_{{\rm PTS}}$ \cite{montanari:06, yaida:16} and according to the RFOT
\cite{kirkpatrick:89,Lubchenko:07,kirkpatrick:15,dzero:09,biroli:09}   $\xi_{{\rm  PTS}}$   diverges   at  the  Kauzmann
temperature  $T_K$ \cite{kauzmann:48},  the temperature  at which  the
configurational entropy per particle $s_c$ is supposed to vanish.  Below $T_K$ the
glass is hypothesized to enter   the so-called \lq \lq  ideal glass state"
\cite{kirkpatrick:89,Lubchenko:07,  biroli:09} where the configurational entropy $s_c$  is zero. The growth of $\xi_{{\rm PTS}}$ is driven by the \lq \lq rarefraction'' of states as $T\to T_K$. (Some suspect \cite{tanaka:03,stillinger:01}, that
$s_c$ never actually vanishes.) 
Now  as the temperature is reduced below the glass transition temperature $T_g$,  $\xi_{{\rm PTS}}$ increases \cite{yaida:16a, berthier:18b}.
 In the vicinity of $T_g$, $\xi_{{\rm PTS}}$ is actually small, only a few molecular diameters \cite{yaida:16,yaida:16a,berthier:18b}. As a consequence of new  simulational methods (the swap algorithm \cite{ediger:16,berthier:18b}) it has now become possible to study  $\xi_{{\rm PTS}}$  at temperatures well-below $T_g$ where indeed it becomes somewhat longer. According to RFOT theory it  diverges as $T \to T_K$, but if the discontinuous transition is removed by fluctuations around the mean-field solution and the transition is avoided, the correlation length will not actually diverge, just as was seen in the Migdal-Kadanoff RG procedure \cite {yeo:12, drossel:02}.

A recent attempt to explain the growing length scale can be found in Refs. \cite{tarzia:18a,tarzia:18b}. This work too focusses on the effect of fluctuations on the mean-field solution. One of their scenarios was that the growing length scale was in the same universality class of the Ising spin glass in a field (a scenario favored by us \cite{moore:06}), but their chief scenario was that there was a genuine RFOT like transition at $T_K$.

 Our work in this paper suggests that the discontinuous transition from the RS paramagnetic state to a state with 1RSB, which was first derived in the fully connected model, might be modified  even in very high dimensions, or just by finite size effects alone, to  a discontinuous transition  to a state with FRSB. As the dimensionality is lowered further, possibly below six dimensions, these same fluctuations might also remove the state with FRSB completely. But our main message is that the nature of the ideal glass state might not be as simple as that envisaged on the basis of it being a state whose order parameter is that predicted by 1RSB, and that much further work remains to be done.

\begin{acknowledgments}

We would like to thank Professor V\'{a}clav Jani\v{s} for information on replica symmetry breaking in Potts spin glass models.
JY was supported by
Basic Science Research Program through the National Research Foundation 
of Korea (NRF) funded by the Ministry of Education (2017R1D1A09000527).

\end{acknowledgments}






\begin{thebibliography}{40}%
\makeatletter
\providecommand \@ifxundefined [1]{%
 \@ifx{#1\undefined}
}%
\providecommand \@ifnum [1]{%
 \ifnum #1\expandafter \@firstoftwo
 \else \expandafter \@secondoftwo
 \fi
}%
\providecommand \@ifx [1]{%
 \ifx #1\expandafter \@firstoftwo
 \else \expandafter \@secondoftwo
 \fi
}%
\providecommand \natexlab [1]{#1}%
\providecommand \enquote  [1]{``#1''}%
\providecommand \bibnamefont  [1]{#1}%
\providecommand \bibfnamefont [1]{#1}%
\providecommand \citenamefont [1]{#1}%
\providecommand \href@noop [0]{\@secondoftwo}%
\providecommand \href [0]{\begingroup \@sanitize@url \@href}%
\providecommand \@href[1]{\@@startlink{#1}\@@href}%
\providecommand \@@href[1]{\endgroup#1\@@endlink}%
\providecommand \@sanitize@url [0]{\catcode `\\12\catcode `\$12\catcode
  `\&12\catcode `\#12\catcode `\^12\catcode `\_12\catcode `\%12\relax}%
\providecommand \@@startlink[1]{}%
\providecommand \@@endlink[0]{}%
\providecommand \url  [0]{\begingroup\@sanitize@url \@url }%
\providecommand \@url [1]{\endgroup\@href {#1}{\urlprefix }}%
\providecommand \urlprefix  [0]{URL }%
\providecommand \Eprint [0]{\href }%
\providecommand \doibase [0]{http://dx.doi.org/}%
\providecommand \selectlanguage [0]{\@gobble}%
\providecommand \bibinfo  [0]{\@secondoftwo}%
\providecommand \bibfield  [0]{\@secondoftwo}%
\providecommand \translation [1]{[#1]}%
\providecommand \BibitemOpen [0]{}%
\providecommand \bibitemStop [0]{}%
\providecommand \bibitemNoStop [0]{.\EOS\space}%
\providecommand \EOS [0]{\spacefactor3000\relax}%
\providecommand \BibitemShut  [1]{\csname bibitem#1\endcsname}%
\let\auto@bib@innerbib\@empty
\bibitem [{\citenamefont {Gardner}(1985)}]{gardner:85}%
  \BibitemOpen
  \bibfield  {author} {\bibinfo {author} {\bibfnamefont {E}~\bibnamefont
  {Gardner}},\ }\bibfield  {title} {\enquote {\bibinfo {title} {Spin glasses
  with $p$-spin interactions},}\ }\href@noop {} {\bibfield  {journal} {\bibinfo
   {journal} {Nuclear Physics B}\ }\textbf {\bibinfo {volume} {257}},\ \bibinfo
  {pages} {747} (\bibinfo {year} {1985})}\BibitemShut {NoStop}%
\bibitem [{\citenamefont {Kirkpatrick}\ \emph {et~al.}(1989)\citenamefont
  {Kirkpatrick}, \citenamefont {Thirumalai},\ and\ \citenamefont
  {Wolynes}}]{kirkpatrick:89}%
  \BibitemOpen
  \bibfield  {author} {\bibinfo {author} {\bibfnamefont {T.~R.}\ \bibnamefont
  {Kirkpatrick}}, \bibinfo {author} {\bibfnamefont {D.}~\bibnamefont
  {Thirumalai}}, \ and\ \bibinfo {author} {\bibfnamefont {P.~G.}\ \bibnamefont
  {Wolynes}},\ }\bibfield  {title} {\enquote {\bibinfo {title} {{Scaling
  concepts for the dynamics of viscous liquids near an ideal glassy state}},}\
  }\href@noop {} {\bibfield  {journal} {\bibinfo  {journal} {Phys. Rev. A}\
  }\textbf {\bibinfo {volume} {40}},\ \bibinfo {pages} {1045--1054} (\bibinfo
  {year} {1989})}\BibitemShut {NoStop}%
\bibitem [{\citenamefont {Lubchenko}\ and\ \citenamefont
  {Wolynes}(2007)}]{Lubchenko:07}%
  \BibitemOpen
  \bibfield  {author} {\bibinfo {author} {\bibfnamefont {Vassiliy}\
  \bibnamefont {Lubchenko}}\ and\ \bibinfo {author} {\bibfnamefont {Peter~G.}\
  \bibnamefont {Wolynes}},\ }\bibfield  {title} {\enquote {\bibinfo {title}
  {{Theory of Structural Glasses and Supercooled Liquids}},}\ }\href@noop {}
  {\bibfield  {journal} {\bibinfo  {journal} {Annual Review of Physical
  Chemistry}\ }\textbf {\bibinfo {volume} {58}},\ \bibinfo {pages} {235--266}
  (\bibinfo {year} {2007})}\BibitemShut {NoStop}%
\bibitem [{\citenamefont {Kirkpatrick}\ and\ \citenamefont
  {Thirumalai}(2015)}]{kirkpatrick:15}%
  \BibitemOpen
  \bibfield  {author} {\bibinfo {author} {\bibfnamefont {T.~R.}\ \bibnamefont
  {Kirkpatrick}}\ and\ \bibinfo {author} {\bibfnamefont {D.}~\bibnamefont
  {Thirumalai}},\ }\bibfield  {title} {\enquote {\bibinfo {title} {{Colloquium:
  Random first order transition theory concepts in biology and physics}},}\
  }\href@noop {} {\bibfield  {journal} {\bibinfo  {journal} {Rev. Mod. Phys.}\
  }\textbf {\bibinfo {volume} {87}},\ \bibinfo {pages} {183--209} (\bibinfo
  {year} {2015})}\BibitemShut {NoStop}%
\bibitem [{\citenamefont {Dzero}\ \emph {et~al.}(2009)\citenamefont {Dzero},
  \citenamefont {Schmalian},\ and\ \citenamefont {Wolynes}}]{dzero:09}%
  \BibitemOpen
  \bibfield  {author} {\bibinfo {author} {\bibfnamefont {Maxim}\ \bibnamefont
  {Dzero}}, \bibinfo {author} {\bibfnamefont {J\"org}\ \bibnamefont
  {Schmalian}}, \ and\ \bibinfo {author} {\bibfnamefont {Peter~G.}\
  \bibnamefont {Wolynes}},\ }\bibfield  {title} {\enquote {\bibinfo {title}
  {Replica theory for fluctuations of the activation barriers in glassy
  systems},}\ }\href@noop {} {\bibfield  {journal} {\bibinfo  {journal} {Phys.
  Rev. B}\ }\textbf {\bibinfo {volume} {80}},\ \bibinfo {pages} {024204}
  (\bibinfo {year} {2009})}\BibitemShut {NoStop}%
\bibitem [{\citenamefont {Biroli}\ and\ \citenamefont
  {Bouchaud}(2010)}]{biroli:09}%
  \BibitemOpen
  \bibfield  {author} {\bibinfo {author} {\bibfnamefont {G.}~\bibnamefont
  {Biroli}}\ and\ \bibinfo {author} {\bibfnamefont {J.~P.}\ \bibnamefont
  {Bouchaud}},\ }\href@noop {} {\emph {\bibinfo {title} {{Structural Glasses
  and Supercooled Liquids:Theory, Experiment,and Applications}}}}\ (\bibinfo
  {publisher} {Wiley, Singapore},\ \bibinfo {year} {2010})\BibitemShut
  {NoStop}%
\bibitem [{\citenamefont {Cavagna}(2009)}]{cavagna:09}%
  \BibitemOpen
  \bibfield  {author} {\bibinfo {author} {\bibfnamefont {Andrea}\ \bibnamefont
  {Cavagna}},\ }\bibfield  {title} {\enquote {\bibinfo {title} {{{Supercooled
  liquids for pedestrians}}},}\ }\href@noop {} {\bibfield  {journal} {\bibinfo
  {journal} {Physics Reports}\ }\textbf {\bibinfo {volume} {476}},\ \bibinfo
  {pages} {51} (\bibinfo {year} {2009})}\BibitemShut {NoStop}%
\bibitem [{\citenamefont {Kauzmann}(1948)}]{kauzmann:48}%
  \BibitemOpen
  \bibfield  {author} {\bibinfo {author} {\bibfnamefont {W.}~\bibnamefont
  {Kauzmann}},\ }\bibfield  {title} {\enquote {\bibinfo {title} {The nature of
  the glassy state and the behavior of liquids at low temperatures},}\
  }\href@noop {} {\bibfield  {journal} {\bibinfo  {journal} {Chem. Reviews}\
  }\textbf {\bibinfo {volume} {43}},\ \bibinfo {pages} {219} (\bibinfo {year}
  {1948})}\BibitemShut {NoStop}%
\bibitem [{\citenamefont {Berthier}\ and\ \citenamefont
  {Ediger}(2016)}]{ediger:16}%
  \BibitemOpen
  \bibfield  {author} {\bibinfo {author} {\bibfnamefont {Ludovic}\ \bibnamefont
  {Berthier}}\ and\ \bibinfo {author} {\bibfnamefont {Mark~D.}\ \bibnamefont
  {Ediger}},\ }\bibfield  {title} {\enquote {\bibinfo {title} {Facets of glass
  physics},}\ }\href@noop {} {\bibfield  {journal} {\bibinfo  {journal}
  {Physics Today}\ }\textbf {\bibinfo {volume} {69}},\ \bibinfo {pages} {40}
  (\bibinfo {year} {2016})}\BibitemShut {NoStop}%
\bibitem [{\citenamefont {Royall}\ \emph {et~al.}(2018)\citenamefont {Royall},
  \citenamefont {Turci}, \citenamefont {Tatsumi}, \citenamefont {Russo},\ and\
  \citenamefont {Robinson}}]{Royall:18}%
  \BibitemOpen
  \bibfield  {author} {\bibinfo {author} {\bibfnamefont {C~Patrick}\
  \bibnamefont {Royall}}, \bibinfo {author} {\bibfnamefont {Francesco}\
  \bibnamefont {Turci}}, \bibinfo {author} {\bibfnamefont {Soichi}\
  \bibnamefont {Tatsumi}}, \bibinfo {author} {\bibfnamefont {John}\
  \bibnamefont {Russo}}, \ and\ \bibinfo {author} {\bibfnamefont {Joshua}\
  \bibnamefont {Robinson}},\ }\bibfield  {title} {\enquote {\bibinfo {title}
  {{The race to the bottom: approaching the ideal glass?}}}\ }\href@noop {}
  {\bibfield  {journal} {\bibinfo  {journal} {Journal of Physics: Condensed
  Matter}\ }\textbf {\bibinfo {volume} {30}},\ \bibinfo {pages} {363001}
  (\bibinfo {year} {2018})}\BibitemShut {NoStop}%
\bibitem [{\citenamefont {Gross}\ \emph {et~al.}(1985)\citenamefont {Gross},
  \citenamefont {Kanter},\ and\ \citenamefont {Sompolinsky}}]{gross:85}%
  \BibitemOpen
  \bibfield  {author} {\bibinfo {author} {\bibfnamefont {D.~J.}\ \bibnamefont
  {Gross}}, \bibinfo {author} {\bibfnamefont {I.}~\bibnamefont {Kanter}}, \
  and\ \bibinfo {author} {\bibfnamefont {H.}~\bibnamefont {Sompolinsky}},\
  }\bibfield  {title} {\enquote {\bibinfo {title} {{Mean-field theory of the
  Potts glass}},}\ }\href@noop {} {\bibfield  {journal} {\bibinfo  {journal}
  {Phys. Rev. Lett.}\ }\textbf {\bibinfo {volume} {55}},\ \bibinfo {pages}
  {304--307} (\bibinfo {year} {1985})}\BibitemShut {NoStop}%
\bibitem [{\citenamefont {{Berthier}}\ \emph {et~al.}(2019)\citenamefont
  {{Berthier}}, \citenamefont {{Biroli}}, \citenamefont {{Charbonneau}},
  \citenamefont {{Corwin}}, \citenamefont {{Franz}},\ and\ \citenamefont
  {{Zamponi}}}]{berthier:19}%
  \BibitemOpen
  \bibfield  {author} {\bibinfo {author} {\bibfnamefont {Ludovic}\ \bibnamefont
  {{Berthier}}}, \bibinfo {author} {\bibfnamefont {Giulio}\ \bibnamefont
  {{Biroli}}}, \bibinfo {author} {\bibfnamefont {Patrick}\ \bibnamefont
  {{Charbonneau}}}, \bibinfo {author} {\bibfnamefont {Eric~I.}\ \bibnamefont
  {{Corwin}}}, \bibinfo {author} {\bibfnamefont {Silvio}\ \bibnamefont
  {{Franz}}}, \ and\ \bibinfo {author} {\bibfnamefont {Francesco}\ \bibnamefont
  {{Zamponi}}},\ }\bibfield  {title} {\enquote {\bibinfo {title} {{Perspective:
  Gardner Physics in Amorphous Solids and Beyond}},}\ }\href@noop {} {\bibfield
   {journal} {\bibinfo  {journal} {arXiv e-prints}\ ,\ \bibinfo {eid}
  {arXiv:1902.10494}} (\bibinfo {year} {2019})},\ \Eprint
  {http://arxiv.org/abs/1902.10494} {arXiv:1902.10494} \BibitemShut {NoStop}%
\bibitem [{\citenamefont {Montanari}\ and\ \citenamefont
  {Ricci-Tersenghi}(2003)}]{montanari:03}%
  \BibitemOpen
  \bibfield  {author} {\bibinfo {author} {\bibfnamefont {A.}~\bibnamefont
  {Montanari}}\ and\ \bibinfo {author} {\bibfnamefont {F.}~\bibnamefont
  {Ricci-Tersenghi}},\ }\bibfield  {title} {\enquote {\bibinfo {title} {On the
  nature of the low-temperature phase in discontinuous mean-field spin
  glasses},}\ }\href@noop {} {\bibfield  {journal} {\bibinfo  {journal} {The
  European Physical Journal B - Condensed Matter and Complex Systems}\ }\textbf
  {\bibinfo {volume} {33}},\ \bibinfo {pages} {339} (\bibinfo {year}
  {2003})}\BibitemShut {NoStop}%
\bibitem [{\citenamefont {Billoire}\ \emph {et~al.}(2005)\citenamefont
  {Billoire}, \citenamefont {Giomi},\ and\ \citenamefont
  {Marinari}}]{billoire:05}%
  \BibitemOpen
  \bibfield  {author} {\bibinfo {author} {\bibfnamefont {A}~\bibnamefont
  {Billoire}}, \bibinfo {author} {\bibfnamefont {L}~\bibnamefont {Giomi}}, \
  and\ \bibinfo {author} {\bibfnamefont {E}~\bibnamefont {Marinari}},\
  }\bibfield  {title} {\enquote {\bibinfo {title} {{The mean-field infinite
  range $p= 3$ spin glass: Equilibrium landscape and correlation time
  scales}},}\ }\href@noop {} {\bibfield  {journal} {\bibinfo  {journal}
  {Europhysics Letters ({EPL})}\ }\textbf {\bibinfo {volume} {71}},\ \bibinfo
  {pages} {824} (\bibinfo {year} {2005})}\BibitemShut {NoStop}%
\bibitem [{\citenamefont {Derrida}\ and\ \citenamefont
  {Mottishaw}(2018)}]{mottishaw:18}%
  \BibitemOpen
  \bibfield  {author} {\bibinfo {author} {\bibfnamefont {Bernard}\ \bibnamefont
  {Derrida}}\ and\ \bibinfo {author} {\bibfnamefont {Peter}\ \bibnamefont
  {Mottishaw}},\ }\bibfield  {title} {\enquote {\bibinfo {title} {{Finite Size
  Corrections to the Parisi Overlap Function in the GREM"}},}\ }\href@noop {}
  {\bibfield  {journal} {\bibinfo  {journal} {Journal of Statistical Physics}\
  }\textbf {\bibinfo {volume} {172}},\ \bibinfo {pages} {592} (\bibinfo {year}
  {2018})}\BibitemShut {NoStop}%
\bibitem [{\citenamefont {Campellone}\ \emph {et~al.}(1999)\citenamefont
  {Campellone}, \citenamefont {Parisi},\ and\ \citenamefont
  {Ranieri}}]{campellone:99}%
  \BibitemOpen
  \bibfield  {author} {\bibinfo {author} {\bibfnamefont {Matteo}\ \bibnamefont
  {Campellone}}, \bibinfo {author} {\bibfnamefont {Giorgio}\ \bibnamefont
  {Parisi}}, \ and\ \bibinfo {author} {\bibfnamefont {Paola}\ \bibnamefont
  {Ranieri}},\ }\bibfield  {title} {\enquote {\bibinfo {title}
  {{Finite-dimensional corrections to the mean field in a short-range $p$-spin
  glassy model}},}\ }\href@noop {} {\bibfield  {journal} {\bibinfo  {journal}
  {Phys. Rev. B}\ }\textbf {\bibinfo {volume} {59}},\ \bibinfo {pages}
  {1036--1045} (\bibinfo {year} {1999})}\BibitemShut {NoStop}%
\bibitem [{\citenamefont {Moore}(2006)}]{moore:06b}%
  \BibitemOpen
  \bibfield  {author} {\bibinfo {author} {\bibfnamefont {M.~A.}\ \bibnamefont
  {Moore}},\ }\bibfield  {title} {\enquote {\bibinfo {title} {{Interface Free
  Energies in $p$-Spin Glass Models}},}\ }\href@noop {} {\bibfield  {journal}
  {\bibinfo  {journal} {Phys. Rev. Lett.}\ }\textbf {\bibinfo {volume} {96}},\
  \bibinfo {pages} {137202} (\bibinfo {year} {2006})}\BibitemShut {NoStop}%
\bibitem [{\citenamefont {{de Dominicis}}\ \emph {et~al.}(1997)\citenamefont
  {{de Dominicis}}, \citenamefont {{Kondor}},\ and\ \citenamefont
  {{Temesv{\'a}ri}}}]{dominicis:97}%
  \BibitemOpen
  \bibfield  {author} {\bibinfo {author} {\bibfnamefont {C.}~\bibnamefont {{de
  Dominicis}}}, \bibinfo {author} {\bibfnamefont {I.}~\bibnamefont {{Kondor}}},
  \ and\ \bibinfo {author} {\bibfnamefont {T.}~\bibnamefont
  {{Temesv{\'a}ri}}},\ }\bibfield  {title} {\enquote {\bibinfo {title} {{Beyond
  the Sherrington-Kirkpatrick Model}},}\ }\href {\doibase
  10.1142/9789812819437_0005} {\bibfield  {journal} {\bibinfo  {journal} {Spin
  Glasses And Random Fields.~Series: Series on Directions in Condensed Matter
  Physics, ISBN: 978-981-02-3183-5,~WORLD SCIENTIFIC, Edited by A P Young,
  vol.~12, pp.~119-160}\ }\textbf {\bibinfo {volume} {12}},\ \bibinfo {pages}
  {119--160} (\bibinfo {year} {1997})},\ \Eprint
  {http://arxiv.org/abs/cond-mat/9705215} {cond-mat/9705215} \BibitemShut
  {NoStop}%
\bibitem [{\citenamefont {Campellone}\ \emph {et~al.}(1998)\citenamefont
  {Campellone}, \citenamefont {Coluzzi},\ and\ \citenamefont
  {Parisi}}]{campellone:98}%
  \BibitemOpen
  \bibfield  {author} {\bibinfo {author} {\bibfnamefont {Matteo}\ \bibnamefont
  {Campellone}}, \bibinfo {author} {\bibfnamefont {Barbara}\ \bibnamefont
  {Coluzzi}}, \ and\ \bibinfo {author} {\bibfnamefont {Giorgio}\ \bibnamefont
  {Parisi}},\ }\bibfield  {title} {\enquote {\bibinfo {title} {{Numerical study
  of a short-range $p$-spin glass model in three dimensions}},}\ }\href
  {https://link.aps.org/doi/10.1103/PhysRevB.58.12081} {\bibfield  {journal}
  {\bibinfo  {journal} {Phys. Rev. B}\ }\textbf {\bibinfo {volume} {58}},\
  \bibinfo {pages} {12081} (\bibinfo {year} {1998})}\BibitemShut {NoStop}%
\bibitem [{\citenamefont {Franz}\ and\ \citenamefont
  {Parisi}(1999)}]{franz:99}%
  \BibitemOpen
  \bibfield  {author} {\bibinfo {author} {\bibfnamefont {S.}~\bibnamefont
  {Franz}}\ and\ \bibinfo {author} {\bibfnamefont {G.}~\bibnamefont {Parisi}},\
  }\bibfield  {title} {\enquote {\bibinfo {title} {Critical properties of a
  three-dimensional p-spin model},}\ }\href@noop {} {\bibfield  {journal}
  {\bibinfo  {journal} {The European Physical Journal B - Condensed Matter and
  Complex Systems}\ }\textbf {\bibinfo {volume} {8}},\ \bibinfo {pages} {417}
  (\bibinfo {year} {1999})}\BibitemShut {NoStop}%
\bibitem [{\citenamefont {{J. Kurchan}}\ \emph {et~al.}(1993)\citenamefont {{J.
  Kurchan}}, \citenamefont {{G.Parisi}},\ and\ \citenamefont {{M.A.
  Virasoro}}}]{kurchan:93}%
  \BibitemOpen
  \bibfield  {author} {\bibinfo {author} {\bibnamefont {{J. Kurchan}}},
  \bibinfo {author} {\bibnamefont {{G.Parisi}}}, \ and\ \bibinfo {author}
  {\bibnamefont {{M.A. Virasoro}}},\ }\bibfield  {title} {\enquote {\bibinfo
  {title} {{Barriers and metastable states as saddle points in the replica
  approach}},}\ }\href@noop {} {\bibfield  {journal} {\bibinfo  {journal} {J.
  Phys. I France}\ }\textbf {\bibinfo {volume} {3}},\ \bibinfo {pages}
  {1819--1838} (\bibinfo {year} {1993})}\BibitemShut {NoStop}%
\bibitem [{\citenamefont {Baviera}\ and\ \citenamefont
  {Virasoro}(2015)}]{Baviera:15}%
  \BibitemOpen
  \bibfield  {author} {\bibinfo {author} {\bibfnamefont {R}~\bibnamefont
  {Baviera}}\ and\ \bibinfo {author} {\bibfnamefont {M~A}\ \bibnamefont
  {Virasoro}},\ }\bibfield  {title} {\enquote {\bibinfo {title} {{A method that
  reveals the multi-level ultrametric tree hidden in
  p{\hspace{0.167em}}-spin-glass-like systems}},}\ }\href@noop {} {\bibfield
  {journal} {\bibinfo  {journal} {{Journal of Statistical Mechanics: Theory and
  Experiment}}\ }\textbf {\bibinfo {volume} {2015}},\ \bibinfo {pages} {P12007}
  (\bibinfo {year} {2015})}\BibitemShut {NoStop}%
\bibitem [{\citenamefont {Moore}\ and\ \citenamefont
  {Drossel}(2002)}]{drossel:02}%
  \BibitemOpen
  \bibfield  {author} {\bibinfo {author} {\bibfnamefont {M.~A.}\ \bibnamefont
  {Moore}}\ and\ \bibinfo {author} {\bibfnamefont {Barbara}\ \bibnamefont
  {Drossel}},\ }\bibfield  {title} {\enquote {\bibinfo {title} {{$p$-Spin Model
  in Finite Dimensions and Its Relation to Structural Glasses}},}\ }\href@noop
  {} {\bibfield  {journal} {\bibinfo  {journal} {Phys. Rev. Lett.}\ }\textbf
  {\bibinfo {volume} {89}},\ \bibinfo {pages} {217202} (\bibinfo {year}
  {2002})}\BibitemShut {NoStop}%
\bibitem [{\citenamefont {Moore}\ and\ \citenamefont {Yeo}(2006)}]{moore:06}%
  \BibitemOpen
  \bibfield  {author} {\bibinfo {author} {\bibfnamefont {M.~A.}\ \bibnamefont
  {Moore}}\ and\ \bibinfo {author} {\bibfnamefont {J.}~\bibnamefont {Yeo}},\
  }\bibfield  {title} {\enquote {\bibinfo {title} {Thermodynamic glass
  transition in finite dimensions},}\ }\href@noop {} {\bibfield  {journal}
  {\bibinfo  {journal} {Phys. Rev. Lett.}\ }\textbf {\bibinfo {volume} {96}},\
  \bibinfo {pages} {095701} (\bibinfo {year} {2006})}\BibitemShut {NoStop}%
\bibitem [{\citenamefont {Caltagirone}\ \emph {et~al.}(2011)\citenamefont
  {Caltagirone}, \citenamefont {Ferrari}, \citenamefont {Leuzzi}, \citenamefont
  {Parisi},\ and\ \citenamefont {Rizzo}}]{caltagirone:11}%
  \BibitemOpen
  \bibfield  {author} {\bibinfo {author} {\bibfnamefont {F.}~\bibnamefont
  {Caltagirone}}, \bibinfo {author} {\bibfnamefont {U.}~\bibnamefont
  {Ferrari}}, \bibinfo {author} {\bibfnamefont {L.}~\bibnamefont {Leuzzi}},
  \bibinfo {author} {\bibfnamefont {G.}~\bibnamefont {Parisi}}, \ and\ \bibinfo
  {author} {\bibfnamefont {T.}~\bibnamefont {Rizzo}},\ }\bibfield  {title}
  {\enquote {\bibinfo {title} {{Ising $M$-$p$-spin mean-field model for the
  structural glass: Continuous versus discontinuous transition}},}\ }\href@noop
  {} {\bibfield  {journal} {\bibinfo  {journal} {Phys. Rev. B}\ }\textbf
  {\bibinfo {volume} {83}},\ \bibinfo {pages} {104202} (\bibinfo {year}
  {2011})}\BibitemShut {NoStop}%
\bibitem [{\citenamefont {Yeo}\ and\ \citenamefont {Moore}(2012)}]{yeo:12}%
  \BibitemOpen
  \bibfield  {author} {\bibinfo {author} {\bibfnamefont {Joonhyun}\
  \bibnamefont {Yeo}}\ and\ \bibinfo {author} {\bibfnamefont {M.~A.}\
  \bibnamefont {Moore}},\ }\bibfield  {title} {\enquote {\bibinfo {title}
  {{Origin of the growing length scale in $M$-$p$-spin glass models}},}\
  }\href@noop {} {\bibfield  {journal} {\bibinfo  {journal} {Phys. Rev. E}\
  }\textbf {\bibinfo {volume} {86}},\ \bibinfo {pages} {052501} (\bibinfo
  {year} {2012})}\BibitemShut {NoStop}%
\bibitem [{\citenamefont {Aspelmeier}\ \emph {et~al.}(2003)\citenamefont
  {Aspelmeier}, \citenamefont {Moore},\ and\ \citenamefont {Young}}]{timo:03}%
  \BibitemOpen
  \bibfield  {author} {\bibinfo {author} {\bibfnamefont {T.}~\bibnamefont
  {Aspelmeier}}, \bibinfo {author} {\bibfnamefont {M.~A.}\ \bibnamefont
  {Moore}}, \ and\ \bibinfo {author} {\bibfnamefont {A.~P.}\ \bibnamefont
  {Young}},\ }\bibfield  {title} {\enquote {\bibinfo {title} {{Interface
  Energies in Ising Spin Glasses}},}\ }\href@noop {} {\bibfield  {journal}
  {\bibinfo  {journal} {Phys. Rev. Lett.}\ }\textbf {\bibinfo {volume} {90}},\
  \bibinfo {pages} {127202} (\bibinfo {year} {2003})}\BibitemShut {NoStop}%
\bibitem [{\citenamefont {Nieuwenhuizen}(1996)}]{Nieuwenhuizen:96}%
  \BibitemOpen
  \bibfield  {author} {\bibinfo {author} {\bibfnamefont {T~M}\ \bibnamefont
  {Nieuwenhuizen}},\ }\bibfield  {title} {\enquote {\bibinfo {title} {A puzzle
  on fluctuations of weights in spin glasses},}\ }\href@noop {} {\bibfield
  {journal} {\bibinfo  {journal} {J. Phys. I France}\ }\textbf {\bibinfo
  {volume} {6}},\ \bibinfo {pages} {109} (\bibinfo {year} {1996})}\BibitemShut
  {NoStop}%
\bibitem [{\citenamefont {Ferrero}\ \emph {et~al.}(1996)\citenamefont
  {Ferrero}, \citenamefont {Parisi},\ and\ \citenamefont
  {Ranieri}}]{Ferrero:96}%
  \BibitemOpen
  \bibfield  {author} {\bibinfo {author} {\bibfnamefont {M~E}\ \bibnamefont
  {Ferrero}}, \bibinfo {author} {\bibfnamefont {G}~\bibnamefont {Parisi}}, \
  and\ \bibinfo {author} {\bibfnamefont {P}~\bibnamefont {Ranieri}},\
  }\bibfield  {title} {\enquote {\bibinfo {title} {Fluctuations in a spin-glass
  model with one replica symmetry breaking},}\ }\href@noop {} {\bibfield
  {journal} {\bibinfo  {journal} {Journal of Physics A: Mathematical and
  General}\ }\textbf {\bibinfo {volume} {29}},\ \bibinfo {pages} {L569}
  (\bibinfo {year} {1996})}\BibitemShut {NoStop}%
\bibitem [{\citenamefont {Wang}\ \emph {et~al.}(2018)\citenamefont {Wang},
  \citenamefont {Moore},\ and\ \citenamefont {Katzgraber}}]{wang:18}%
  \BibitemOpen
  \bibfield  {author} {\bibinfo {author} {\bibfnamefont {Wenlong}\ \bibnamefont
  {Wang}}, \bibinfo {author} {\bibfnamefont {M.~A.}\ \bibnamefont {Moore}}, \
  and\ \bibinfo {author} {\bibfnamefont {Helmut~G.}\ \bibnamefont
  {Katzgraber}},\ }\bibfield  {title} {\enquote {\bibinfo {title} {{Fractal
  dimension of interfaces in Edwards-Anderson spin glasses for up to six space
  dimensions}},}\ }\href@noop {} {\bibfield  {journal} {\bibinfo  {journal}
  {Phys. Rev. E}\ }\textbf {\bibinfo {volume} {97}},\ \bibinfo {pages} {032104}
  (\bibinfo {year} {2018})}\BibitemShut {NoStop}%
\bibitem [{\citenamefont {Wang}\ \emph {et~al.}(2017)\citenamefont {Wang},
  \citenamefont {Moore},\ and\ \citenamefont {Katzgraber}}]{wang:17}%
  \BibitemOpen
  \bibfield  {author} {\bibinfo {author} {\bibfnamefont {Wenlong}\ \bibnamefont
  {Wang}}, \bibinfo {author} {\bibfnamefont {M.~A.}\ \bibnamefont {Moore}}, \
  and\ \bibinfo {author} {\bibfnamefont {Helmut~G.}\ \bibnamefont
  {Katzgraber}},\ }\bibfield  {title} {\enquote {\bibinfo {title} {{Fractal
  dimension of interfaces in Edwards-Anderson and long-range Ising spin
  glasses: Determining the applicability of different theoretical
  descriptions}},}\ }\href@noop {} {\bibfield  {journal} {\bibinfo  {journal}
  {Phys. Rev. Lett.}\ }\textbf {\bibinfo {volume} {119}},\ \bibinfo {pages}
  {100602} (\bibinfo {year} {2017})}\BibitemShut {NoStop}%
\bibitem [{\citenamefont {Moore}\ and\ \citenamefont {Bray}(2011)}]{moore:11}%
  \BibitemOpen
  \bibfield  {author} {\bibinfo {author} {\bibfnamefont {M.~A.}\ \bibnamefont
  {Moore}}\ and\ \bibinfo {author} {\bibfnamefont {A.~J.}\ \bibnamefont
  {Bray}},\ }\bibfield  {title} {\enquote {\bibinfo {title} {{Disappearance of
  the de Almeida-Thouless line in six dimensions}},}\ }\href@noop {} {\bibfield
   {journal} {\bibinfo  {journal} {Phys. Rev. B}\ }\textbf {\bibinfo {volume}
  {83}},\ \bibinfo {pages} {224408} (\bibinfo {year} {2011})}\BibitemShut
  {NoStop}%
\bibitem [{\citenamefont {Montanari}\ and\ \citenamefont
  {Semerjian}(2006)}]{montanari:06}%
  \BibitemOpen
  \bibfield  {author} {\bibinfo {author} {\bibfnamefont {Andrea}\ \bibnamefont
  {Montanari}}\ and\ \bibinfo {author} {\bibfnamefont {Guilhem}\ \bibnamefont
  {Semerjian}},\ }\bibfield  {title} {\enquote {\bibinfo {title} {Rigorous
  inequalities between length and time scales in glassy systems},}\ }\href@noop
  {} {\bibfield  {journal} {\bibinfo  {journal} {Journal of Statistical
  Physics}\ }\textbf {\bibinfo {volume} {125}},\ \bibinfo {pages} {23}
  (\bibinfo {year} {2006})}\BibitemShut {NoStop}%
\bibitem [{\citenamefont {Yaida}\ \emph {et~al.}(2016)\citenamefont {Yaida},
  \citenamefont {Berthier}, \citenamefont {Charbonneau},\ and\ \citenamefont
  {Tarjus}}]{yaida:16}%
  \BibitemOpen
  \bibfield  {author} {\bibinfo {author} {\bibfnamefont {Sho}\ \bibnamefont
  {Yaida}}, \bibinfo {author} {\bibfnamefont {Ludovic}\ \bibnamefont
  {Berthier}}, \bibinfo {author} {\bibfnamefont {Patrick}\ \bibnamefont
  {Charbonneau}}, \ and\ \bibinfo {author} {\bibfnamefont {Gilles}\
  \bibnamefont {Tarjus}},\ }\bibfield  {title} {\enquote {\bibinfo {title}
  {Point-to-set lengths, local structure, and glassiness},}\ }\href@noop {}
  {\bibfield  {journal} {\bibinfo  {journal} {Phys. Rev. E}\ }\textbf {\bibinfo
  {volume} {94}},\ \bibinfo {pages} {032605} (\bibinfo {year}
  {2016})}\BibitemShut {NoStop}%
\bibitem [{\citenamefont {Tanaka}(2003)}]{tanaka:03}%
  \BibitemOpen
  \bibfield  {author} {\bibinfo {author} {\bibfnamefont {Hajime}\ \bibnamefont
  {Tanaka}},\ }\bibfield  {title} {\enquote {\bibinfo {title} {{Possible
  resolution of the Kauzmann paradox in supercooled liquids}},}\ }\href@noop {}
  {\bibfield  {journal} {\bibinfo  {journal} {Phys. Rev. E}\ }\textbf {\bibinfo
  {volume} {68}},\ \bibinfo {pages} {011505} (\bibinfo {year}
  {2003})}\BibitemShut {NoStop}%
\bibitem [{\citenamefont {Stillinger}\ \emph {et~al.}(2001)\citenamefont
  {Stillinger}, \citenamefont {Debenedetti},\ and\ \citenamefont
  {Truskett}}]{stillinger:01}%
  \BibitemOpen
  \bibfield  {author} {\bibinfo {author} {\bibfnamefont {Frank~H.}\
  \bibnamefont {Stillinger}}, \bibinfo {author} {\bibfnamefont {Pablo~G.}\
  \bibnamefont {Debenedetti}}, \ and\ \bibinfo {author} {\bibfnamefont
  {Thomas~M.}\ \bibnamefont {Truskett}},\ }\bibfield  {title} {\enquote
  {\bibinfo {title} {{The Kauzmann Paradox Revisited}},}\ }\href@noop {}
  {\bibfield  {journal} {\bibinfo  {journal} {The Journal of Physical Chemistry
  B}\ }\textbf {\bibinfo {volume} {105}},\ \bibinfo {pages} {11809--11816}
  (\bibinfo {year} {2001})}\BibitemShut {NoStop}%
\bibitem [{\citenamefont {Berthier}\ \emph {et~al.}(2016)\citenamefont
  {Berthier}, \citenamefont {Charbonneau},\ and\ \citenamefont
  {Yaida}}]{yaida:16a}%
  \BibitemOpen
  \bibfield  {author} {\bibinfo {author} {\bibfnamefont {Ludovic}\ \bibnamefont
  {Berthier}}, \bibinfo {author} {\bibfnamefont {Patrick}\ \bibnamefont
  {Charbonneau}}, \ and\ \bibinfo {author} {\bibfnamefont {Sho}\ \bibnamefont
  {Yaida}},\ }\bibfield  {title} {\enquote {\bibinfo {title} {Efficient
  measurement of point-to-set correlations and overlap fluctuations in
  glass-forming liquids},}\ }\href@noop {} {\bibfield  {journal} {\bibinfo
  {journal} {The Journal of Chemical Physics}\ }\textbf {\bibinfo {volume}
  {144}},\ \bibinfo {pages} {024501} (\bibinfo {year} {2016})}\BibitemShut
  {NoStop}%
\bibitem [{\citenamefont {Berthier}\ \emph {et~al.}(2019)\citenamefont
  {Berthier}, \citenamefont {Charbonneau}, \citenamefont {Ninarello},
  \citenamefont {Ozawa},\ and\ \citenamefont {Yaida}}]{berthier:18b}%
  \BibitemOpen
  \bibfield  {author} {\bibinfo {author} {\bibfnamefont {Ludovic}\ \bibnamefont
  {Berthier}}, \bibinfo {author} {\bibfnamefont {Patrick}\ \bibnamefont
  {Charbonneau}}, \bibinfo {author} {\bibfnamefont {Andrea}\ \bibnamefont
  {Ninarello}}, \bibinfo {author} {\bibfnamefont {Misaki}\ \bibnamefont
  {Ozawa}}, \ and\ \bibinfo {author} {\bibfnamefont {Sho}\ \bibnamefont
  {Yaida}},\ }\bibfield  {title} {\enquote {\bibinfo {title} {Zero-temperature
  glass transition in two dimensions},}\ }\href@noop {} {\bibfield  {journal}
  {\bibinfo  {journal} {{Nature Communications}}\ }\textbf {\bibinfo {volume}
  {10}},\ \bibinfo {pages} {1508} (\bibinfo {year} {2019})}\BibitemShut
  {NoStop}%
\bibitem [{\citenamefont {Biroli}\ \emph
  {et~al.}(2018{\natexlab{a}})\citenamefont {Biroli}, \citenamefont
  {Cammarota}, \citenamefont {Tarjus},\ and\ \citenamefont
  {Tarzia}}]{tarzia:18a}%
  \BibitemOpen
  \bibfield  {author} {\bibinfo {author} {\bibfnamefont {Giulio}\ \bibnamefont
  {Biroli}}, \bibinfo {author} {\bibfnamefont {Chiara}\ \bibnamefont
  {Cammarota}}, \bibinfo {author} {\bibfnamefont {Gilles}\ \bibnamefont
  {Tarjus}}, \ and\ \bibinfo {author} {\bibfnamefont {Marco}\ \bibnamefont
  {Tarzia}},\ }\bibfield  {title} {\enquote {\bibinfo {title} {{Random-field
  Ising-like effective theory of the glass transition. I. Mean-field
  models}},}\ }\href@noop {} {\bibfield  {journal} {\bibinfo  {journal} {Phys.
  Rev. B}\ }\textbf {\bibinfo {volume} {98}},\ \bibinfo {pages} {174205}
  (\bibinfo {year} {2018}{\natexlab{a}})}\BibitemShut {NoStop}%
\bibitem [{\citenamefont {Biroli}\ \emph
  {et~al.}(2018{\natexlab{b}})\citenamefont {Biroli}, \citenamefont
  {Cammarota}, \citenamefont {Tarjus},\ and\ \citenamefont
  {Tarzia}}]{tarzia:18b}%
  \BibitemOpen
  \bibfield  {author} {\bibinfo {author} {\bibfnamefont {Giulio}\ \bibnamefont
  {Biroli}}, \bibinfo {author} {\bibfnamefont {Chiara}\ \bibnamefont
  {Cammarota}}, \bibinfo {author} {\bibfnamefont {Gilles}\ \bibnamefont
  {Tarjus}}, \ and\ \bibinfo {author} {\bibfnamefont {Marco}\ \bibnamefont
  {Tarzia}},\ }\bibfield  {title} {\enquote {\bibinfo {title} {{Random field
  Ising-like effective theory of the glass transition. II. Finite-dimensional
  models}},}\ }\href@noop {} {\bibfield  {journal} {\bibinfo  {journal} {Phys.
  Rev. B}\ }\textbf {\bibinfo {volume} {98}},\ \bibinfo {pages} {174206}
  (\bibinfo {year} {2018}{\natexlab{b}})}\BibitemShut {NoStop}%
\end{thebibliography}
%

\end{document}